\documentclass[aps,prl,longbibliography,twocolumn,showpacs,noeprint]{revtex4-2}

\usepackage{CJK}

\usepackage{graphicx}
\usepackage{amsmath}
\usepackage{dcolumn}
\usepackage{balance}

\usepackage{bm}
\usepackage[normalem]{ulem}
\usepackage{color}
\usepackage{hyperref}

\usepackage{epstopdf}
\newcommand{\bew}{\begin{widetext}}
\newcommand{\ew}{\end{widetext}}

\newcommand{\bq}{\mathbf{q}}

\newcommand{\br}{\mathbf{r}}

\newcommand{\sep}{ \ \ \ , \ \ \ }

\newcommand{\beq}{\begin{equation}}
\newcommand{\eeq}{\end{equation}}
\newcommand{\beqn}{\begin{eqnarray}}
\newcommand{\eeqn}{\end{eqnarray}}
\newcommand{\pp}{\partial}
\newcommand{\dd}{{\rm d}}
\newcommand{\ee}{{\rm e}}

\newcommand{\cO}{{\cal O}}

\newcommand{\la}{\langle}

\newcommand{\ra}{\rangle}

\newcommand{\vnab}{{\bf \nabla}}

\allowdisplaybreaks[1]
\usepackage{amsfonts}

\usepackage{placeins}
\usepackage{float}

\begin{document}
\title{Universal behavior at the Lifshitz Points of an active Malthusian Ising model}
\author{Gabriel Legrand}
\author{Chiu Fan Lee}
\email{c.lee@imperial.ac.uk}
\address{Department of Bioengineering, Imperial College London, South Kensington Campus, London SW7 2AZ, U.K.}
\date{\today}

	\begin{abstract}
Lifshitz points (LPs) are multicritical points where ordered, disordered, and patterned phases meet. Originally studied in equilibrium magnetic systems, LPs have since been identified in soft matter and even cosmological settings. Their role in active, living matter, however, remains entirely unexplored.
Here we address this gap by introducing and analyzing LPs in the Active Malthusian Ising Model (AMIM)---a minimal model of living matter that incorporates motility together with birth–death dynamics. Despite its simplicity, the AMIM provides direct experimental relevance. We show that the system generically exhibits two distinct LPs and elucidate their universal behavior using a dynamic renormalization group analysis with the $\epsilon$-expansion method at one loop.
Our results yield testable predictions for future simulations and experiments, establishing LPs as a fertile testing ground for novel physics in active matter.
	
	\end{abstract}

\maketitle

\begin{figure}[t]
    \includegraphics[width=1\linewidth]{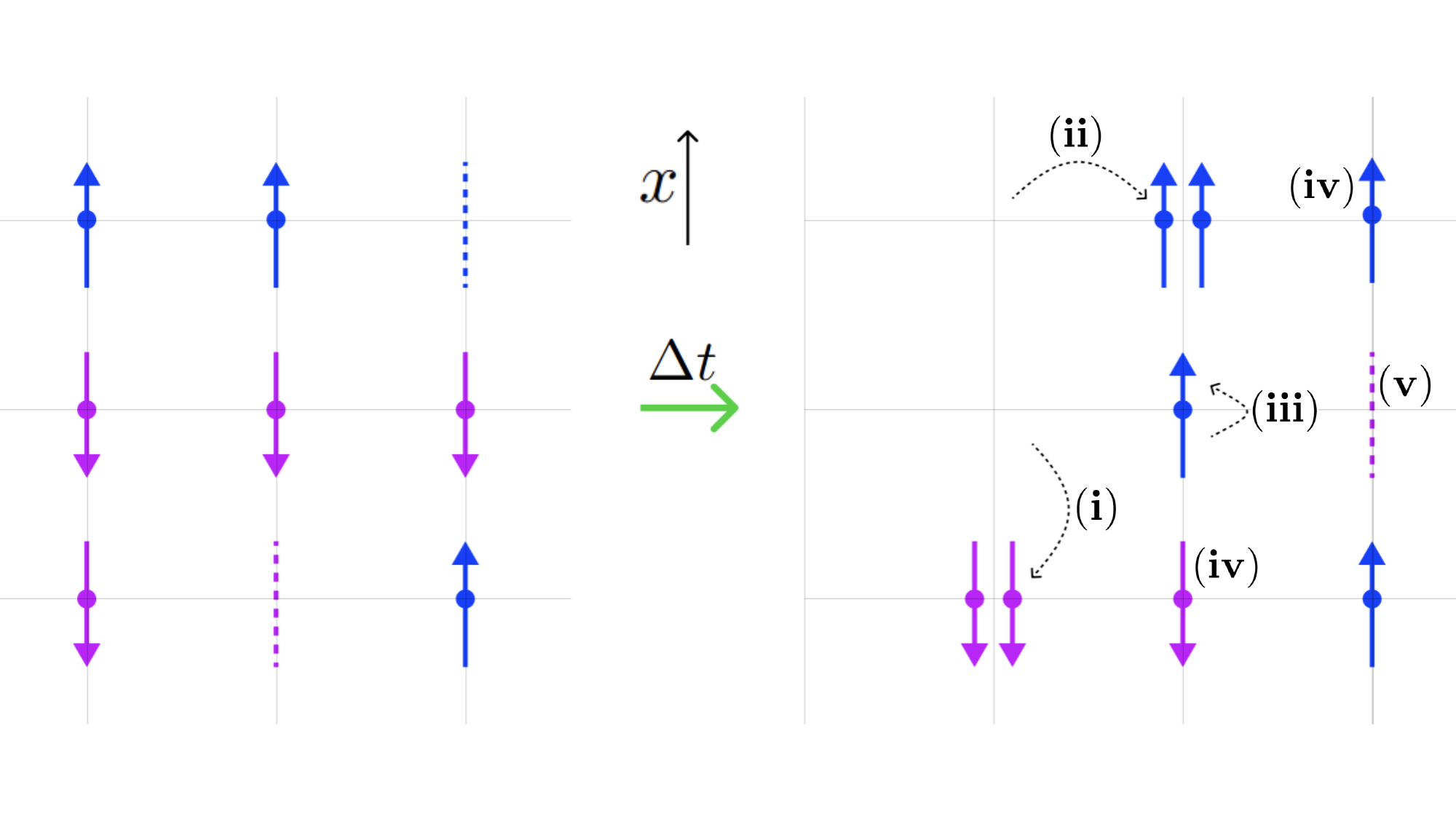}
    \caption{{\it A microscopic active Malthusian Ising model (MAIM).} 
(a) \& (b) In this MAIM, spins' directions preferentially align with the vertical $x$-axis and they dictate the spins' direction motion (i). However, fluctuations can modify spins directions, leading to the spins moving sideways (ii), and spin-flips (iii). Further, we allow for the appearance (or birth) of particle (iv) and disappearance (or death) of particle (v), thus leading to the fact that the particle number is not conserved (Malthusian dynamics).}
    \label{fig}
\end{figure}

In  magnetic systems, Lifshitz points (LPs) are multicritical points where ordered, disordered, and patterned phases meet \cite{hornreich_prl75,hornreich_zpb79}. Besides magnetism, LPs have been identified in diverse physical contexts, from soft matter \cite{michelson_prl77} to quantum gravity \cite{horava_prd09}. Yet, their role beyond inanimate objects remains unexplored. Here, we take a first step toward elucidating the novel physics at the  LPs  in nonequilibrium systems composed of motile constituents \cite{ramaswamy_annrev10,marchetti_rmp13}---a hallmark of animate matter.

{\it Active Malthusian Ising Model (AMIM)---}As a minimal framework, we study the Active Malthusian Ising Model (AMIM). We begin with the Ising model with nonconserved (Model A) dynamics \cite{hohenberg_rmp77}, where the Ising variable is interpreted as the momentum density along a chosen easy axis. To this we add a nonequilibrium advective term representing active motility. If the spin number density were conserved, the resulting system would correspond to the active Ising model \cite{solon_prl13,solon_pre15}. Here, however, we focus on the “Malthusian” version \cite{toner_prl12,chen_prl20,chen_pre20}, in which the spin number density is not conserved due to processes such as birth and death of the motile constituents. A schematic microscopic realization of the AMIM is shown in Fig.~\ref{fig}.

Although minimal, the AMIM is directly relevant to biological contexts. For instance, motile cells migrating through a polymeric gel may experience an intrinsic easy axis induced by stretch-alignment of the gel, while cell reproduction and death naturally give rise to the Malthusian dynamics.

The critical behavior of the AMIM at the Ising transition is known to be governed by the Wilson–Fisher universality class (UC) \cite{bassler_prl94}, whereas the active Ising model with conserved particle number falls into a distinct, nonequilibrium UC \cite{wong_a25}. 
 Here, we instead focus on the multicritical Lifshitz point (LP). The final ingredient required for its realization is the emergence of a patterned state, which arises mathematically from inverting the sign of the Laplacian term in the equation of motion, thereby destabilizing the homogeneous state. Such finite-wavelength instabilities are common in cellular systems undergoing autonomous sorting, often modeled using Cahn–Hilliard-type equations \cite{graner_prl92,glazier_pre93}, and have also been invoked in studies of bacterial swarm dynamics \cite{wensink_pnas12}.

Having motivated the relevance of LPs in the AMIM for experimentally accessible active systems, we now proceed to derive the generic dynamical equations governing this model.

{\it Model equation---}As we interpret the Ising variable, $\phi$, in the AMIM as the momentum density field, the Ising spin direction is naturally coupled to a particular spatial direction. Here, without loss of generality, we choose that direction to be along the $x$ axis. Due to this spin-space coupling, the equilibrium Ising symmetry now becomes the symmetry that respects  the simultaneous inversions:  $x \mapsto -x$, $\phi \mapsto -\phi$. Around the critical point, the mean value of $\phi$ goes continuously through zero, we will therefore  expand the system's model equation in powers of $\phi$, leading to the following generic equation:
\beq
\label{eq:mainA}
\pp_t \phi +\lambda \phi \pp_x \phi = (\mu_x \pp_{x}^2 + \mu_\perp\nabla_\perp^2) \phi -a \phi -b\phi^3+ f
\eeq
where $\la f(\br,t )f(\br',t') \ra = 2D \delta^d(\br-\br')\delta(t-t')$, and we have omitted higher ordered terms that are irrelevant to the leading hydrodynamic behavior. 

Note that the nonequilibrium advective term, $\lambda \phi \pp_x \phi$, appears naturally in this symmetry-based consideration, and this is the term that renders this model distinct from its equilibrium counterpart. Besides the natural emergence of this nonequilibrium term due to the spin-space coupling, the ``diffusion'' coefficients, $\mu$'s, are now also generically distinct depending on their associated spatial dimensions.
To further support the universal nature of the  model equation above, we re-derive the equation from the Malthusian Toner-Tu model with an easy axis in the supplemental material (SM) \cite{SM}.

{\it Two Lifshitz points---}In the equilibrium Ising model under non-conservative dynamics (i.e., when the $\lambda$ term is absent and when $\mu_x =\mu_\perp$), the LP corresponds to fine tuning both $a$ and $\mu_{x,\perp}$ to zero \cite{hornreich_zpb79}. 
Here, since we have two distinct $\mu$'s, there are now generically two distinct LPs: 1) the {\it longitudinal} LP occurs when $\mu_x=0$ and $\mu_\perp>0$, and 2) the {\it transverse} LP occurs when $\mu_x>0$ and $\mu_\perp=0$.
We will now study these two distinct LPs in turn using DRG analyses.

\begin{figure}
         \includegraphics[width=\linewidth]{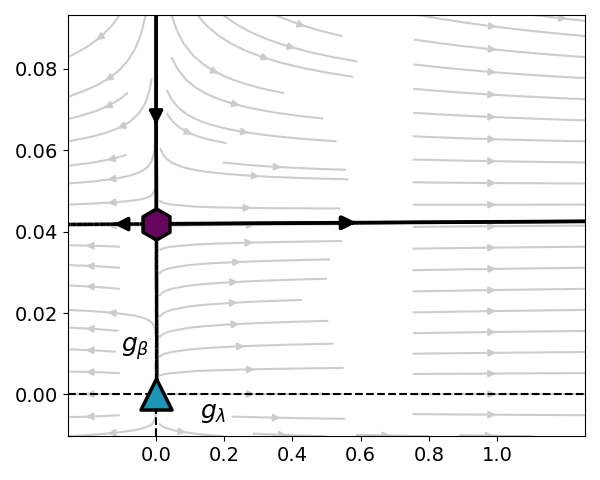}
         \includegraphics[width=\linewidth]{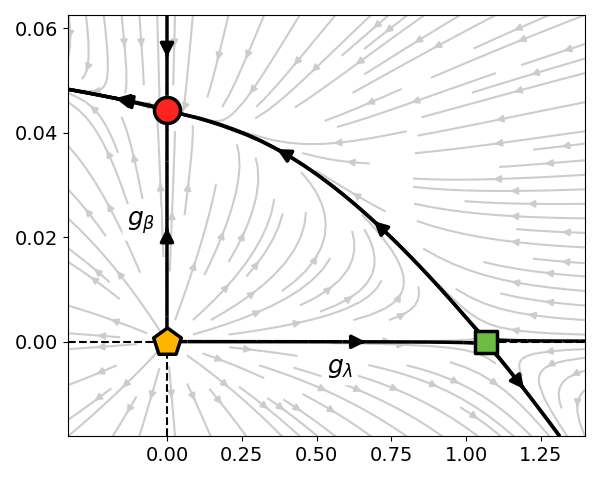}
         \caption{{\it RG flow diagrams of the two distinct Lifshitz Points (LPs)} (a) Longitudinal LP: A generically divergent RG flow is observed for nonzero $\lambda$.     
The RG flow is generated for $d=4.4$, with the 
         Gaussian fixed point (FP) depicted by the blue triangle, and the equilibrium anisotropic Ising LP FP depicted by the purple hexagon.
(b) Transverse LP: The LP multicritical LP behavior is generically described by the equilibrium anisotropic Ising LP FP (red circle). Upon further fine tuning $\beta$ to zero, a new FP emerges (green square). 
The RG flow is generated for $d=6.9$ and the Gaussian FP is depicted by the yellow pentagon.
    }
        \label{fig:RGflow}
 \end{figure}

{\it Longitudinal Lifshitz point (LLP)---}At the bare level, the EOM at this LP is as follows:
\beq
\partial_t\phi+ \frac{\lambda}{2}\partial_x \phi^2 = -\nu_x\partial_x^4\phi + \mu_\bot\vnab_\bot^2\phi    
-\beta \phi^3 +f
 \ ,
 \label{eq:Leom}
\eeq
where we have added the higher ordered $\nu_x\partial_x^4\phi $ term for stability reason (since $\mu_x=0$).
At the linear level, the $\phi$-$\phi$ correlation function can be readily calculated by using the Fourier transform method, leading to the following scaling form at this LLP \cite{SM}:
\beq
 \la \phi({\bf 0},0) \phi(\br,t)\ra = \br^{-2\chi^{\rm lin}_L} S^{\rm lin}_L \left(\frac{t}{\br_\perp^{z^{\rm lin}_L}}, \frac{x}{\br_\perp^{\zeta^{\rm lin}_L}} \right) \ ,
    \eeq
where $S^{\rm lin}_L$ is a universal scaling function at the linear level, and the values of the scaling exponents are:
\beq
\label{eq:Llin}
z^{\rm lin}_L =2 \sep \zeta^{\rm lin}_L =\frac{1}{2} \sep \chi_{L}^{\rm lin} =\frac{5-2d}{4} \ .
\eeq
Using these scaling exponent, one can then readily apply the simple power counting method on the EOM (\ref{eq:Leom}) to ascertain that 1) the upper critical dimension, $d_{L,c}$ is 5.5, and 2) the $\lambda$ term  becomes relevant below $d_{L,c}$, while the $\beta$ term remains irrelevant until $d=4.5$ (based on the linear exponents).

Having identified the relevant nonlinearity, we will now analyze the EOM using the DRG together with the $\epsilon$-perturbation method to the 1-loop level. We leave the details of the analytical calculation in the SM and will just quote the resulting RG flow equations: 
\begin{align}
    \frac{\dd \ln \mu_\perp}{\dd l}&=z-2 \ ,\\
    \frac{\dd \ln \nu_x}{\dd l}&=z_L-4\zeta -\frac{75}{1024\pi\sqrt{2}}g_\lambda^x \ ,\\
    \frac{\dd \ln \lambda}{\dd l}&=\chi+z-\zeta\ , \\
    \frac{\dd \ln D}{\dd l}&=\frac{z-\zeta-2\chi-d+1}{2}\ ,
\end{align}
where
\begin{equation}
    g_\lambda^x=\frac{D\lambda^2}{(\mu_\perp^3\nu_x^9)^\frac{1}{4}}\frac{S_{d-2}\Lambda^{d-5.5}}{(2\pi)^{d-1}}
    \ .
\end{equation}

Translating the RG flow equations into the flow equation for $g_\lambda^x$, we find, with $\epsilon_L=5.5-d$,
\begin{equation}
    \frac{\dd g_\lambda^x}{\dd l}=\epsilon_Lg_\lambda^x +\frac{225}{4096\sqrt{2}}(g_\lambda^x)^2 \ .
\end{equation}
Due to the positive signs in the RG equation above, it clearly indicates a divergent RG flow, at least to this 1-loop level. In the SM, we further test whether adding the $\beta$ nonlinearity, which becomes relevant at $d=4.5$ (based again on the linear exponents (\ref{eq:Llin})), would change this picture using an {\it uncontrolled}, fixed dimension 1-loop calculation. And our conclusion remains the same. The resulting RG flow with the $\beta$ incorporated is shown in Fig.~\ref{fig:RGflow}(a).

Intriguingly, our results can be directly compared with the critical behavior of the Katz–Lebowitz–Spohn (KLS) model of driven lattice gas \cite{katz_prb83,katz_jstatphys84}. In particular, when the $\beta$ term is neglected, the governing equation of the AMIM reduces to a form nearly identical to that of the KLS model at its longitudinal critical point \cite{janssen_ZPB86,leung_jstatphys86}, provided that  the noise is taken to be conservative---as required in the KLS model by particle number conservation. Despite this difference in the noise structure between the KLS model and the AMIM here, both RG analyses  likewise revealed a divergent flow, lending further support to our conclusion that the $\lambda$ nonlinearity generically drives the RG flow to diverge at the LLP of the AMIM.

In general, a divergent RG flow may signal either fluctuation-induced first-order phase separation \cite{kardar_b07} or the existence of a strong-coupling fixed point that controls the scaling behavior \cite{kardar_prl86}. While our perturbative analysis cannot distinguish between these possibilities, it does highlight the exciting need for further investigation using nonperturbative RG techniques \cite{dupuis_physrep21} and  simulations.

\begin{table}[t]
    \begin{tabular}[c]{c|c|c|c|c|c|c}
    \hline
LPs &    FP & Instability & [$g^*_\lambda$,$g^*_\beta$] & $z$ & $\chi$ & $\zeta$ \\
    \hline\hline
L &    \rule{0pt}{3ex} \includegraphics[scale=0.018]{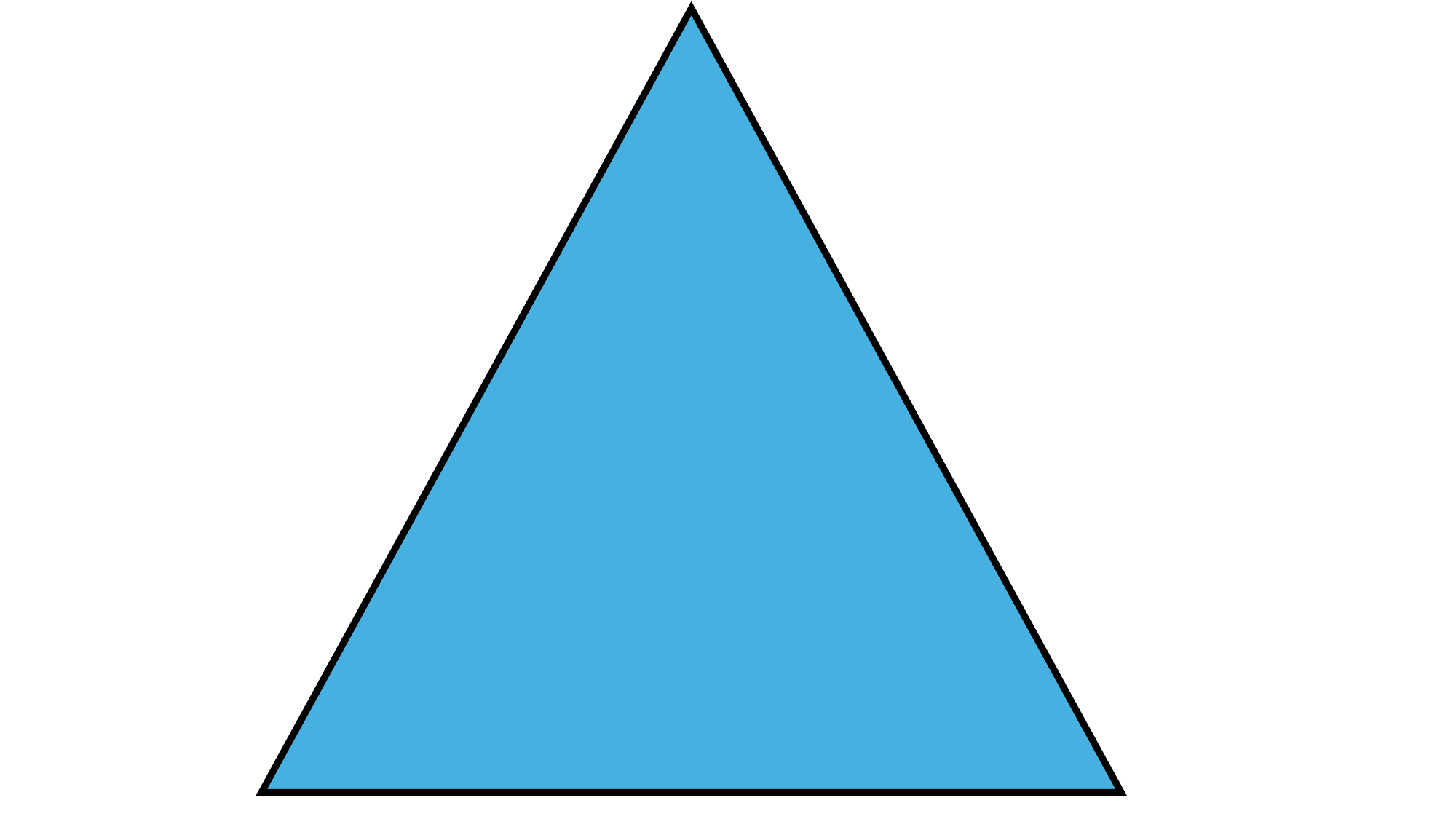} & 2D & $[0,0]$& 2 & $\frac{1}{2}\epsilon_L-\frac{3}{2}$ & $\frac{1}{2}$ \\ [5pt]
L &    \rule{0pt}{3ex}\hspace{2.7pt}\includegraphics[scale=0.018]{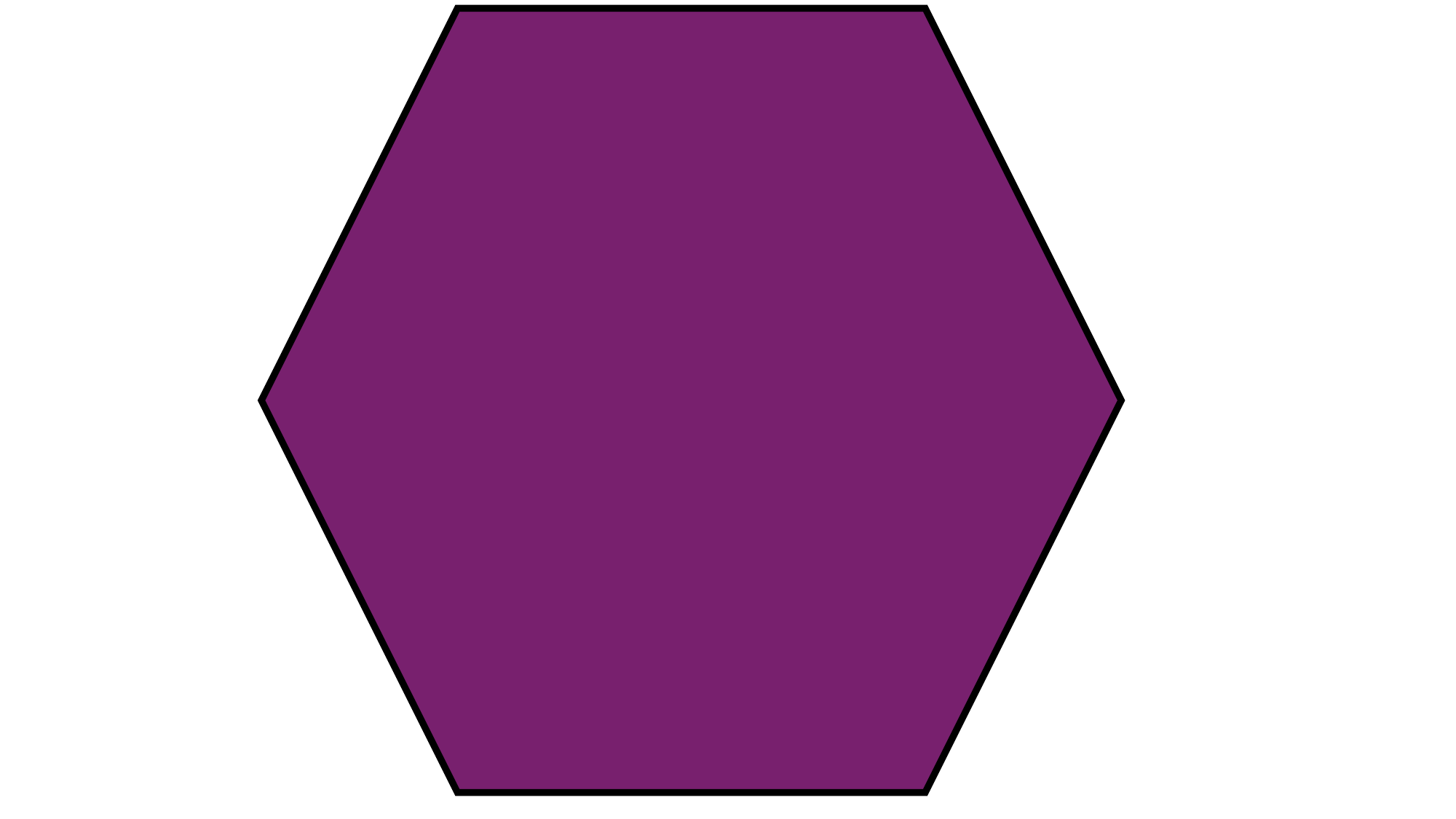} & 1D & $\left[0,\frac{8\sqrt{2}}{27}(\epsilon_L-1)\right]$& 2 & $\frac{1}{2}\epsilon_L-\frac{3}{2}$ & $\frac{1}{2}$ \\ [5pt]
T &    \rule{0pt}{3ex} \hspace{0.8pt}\includegraphics[scale=0.018]{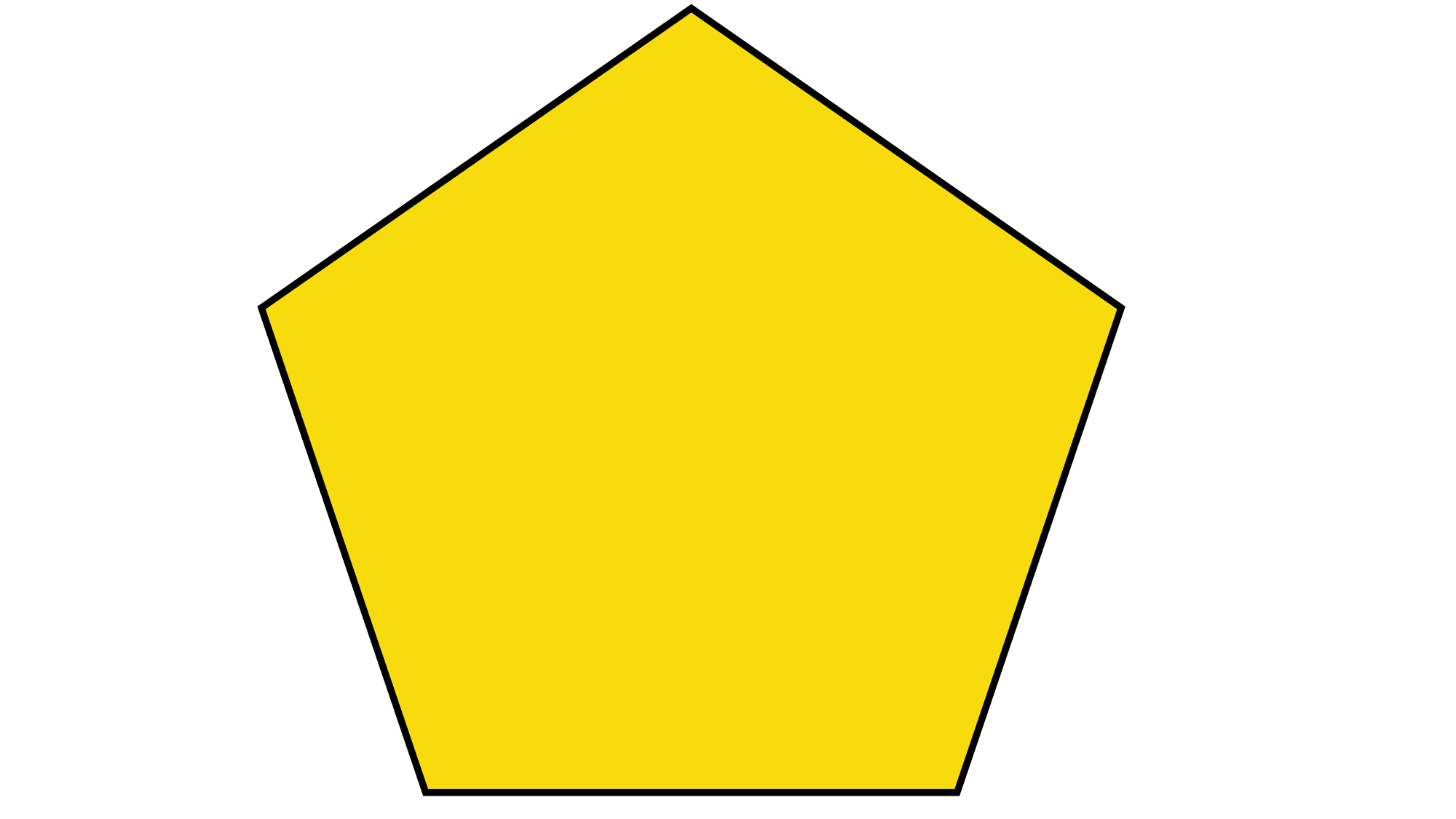} & 2D & $[0,0]$ & 4 & $\frac{1}{2}\epsilon_T-2$ & 2 \\ [5pt]
T &    \rule{0pt}{3ex} \hspace{0.4pt}\includegraphics[scale=0.018]{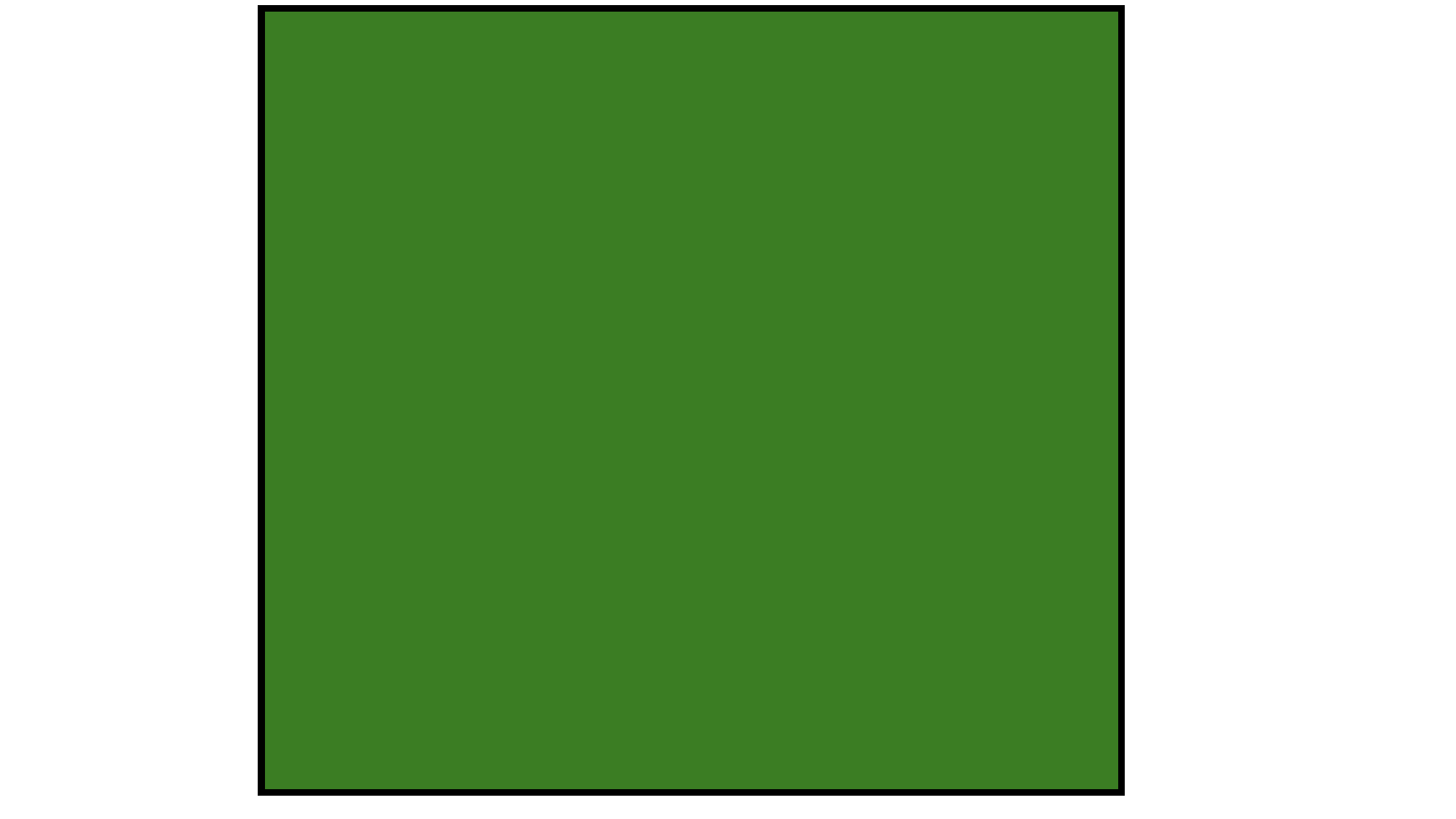} & 1D & $[\frac{32}{3},0]\epsilon_T$& 4 & $\frac{1}{3}\epsilon_T-2$ & $\frac{1}3\epsilon_T+2$\\ [5pt]
T &    \rule{0pt}{3ex} \hspace{1,1pt}\includegraphics[scale=0.018]{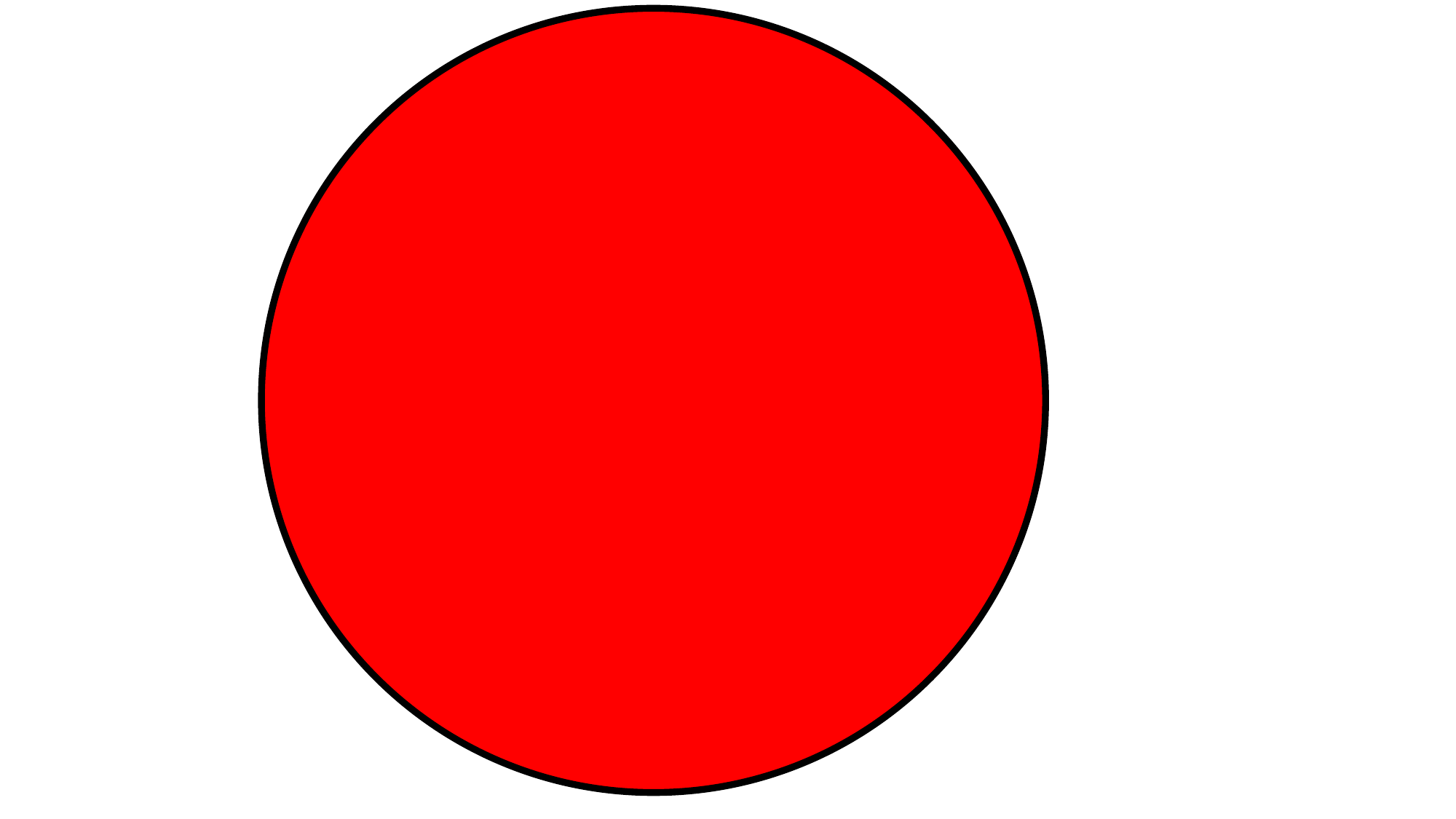} & Stable & $[0,\frac{4}{9}]\epsilon_T$& 4 & $\frac{1}{2}\epsilon_T-2$ & $2$\\ [3pt]
     \hline
    \end{tabular}
    \caption{
{\it Types of LPs, RG FPs, their stabilities, locations, \& critical exponents}. 
Two FPs are found at the longitudinal LPs (L), and three FPs are found at the transverse LPs (T).  The symbols next  used in Fig.~\ref{fig:RGflow} to depict their locations. The Instability column shows the number of unstable direction of each FP within the their respective multicritical manifold ($\alpha = \mu_x =0$ for LLP and $\alpha = \mu_\perp =0$ for TLP). The FP locations are shown by the $g^*$'s. 
 The subsequent 3 columns show the critical exponents. We use $\epsilon_L=5.5-d$ and $\epsilon_T=7-d$.
}
    \label{table}
\end{table}

{\it Transverse Lifshitz point (TLP)---}Here, the EOM at the bare level is as follows:
\beq
\partial_t\phi+ \frac{\lambda}{2}\partial_x \phi^2 = \mu_x\partial_x^2\phi - \nu_\bot\vnab_\bot^4\phi    
-\beta \phi^3 +f
 \ ,
 \label{eq:Leom}
\eeq
where we have added the higher ordered $\nu_\perp\nabla^4\phi$ term again for stability reason. Calculating the $\phi$-$\phi$ correlation function at the linear level as before leads to the following scaling exponents \cite{SM}:
\beq
\label{eq:Llin}
z^{\rm lin}_T =4 \sep \zeta^{\rm lin}_T =2 \sep \chi_{T}^{\rm lin} =\frac{3-d}{2} \ .
\eeq
Using these scaling exponents, we find that 1) the upper critical dimension, $d_{T,c}$ is 7, and 2) both $\lambda$ and $\beta$ nonlinearities become relevant for $d<d_{T,c}$.

At the 1-loop level, the RG flow equations are \cite{SM}:
\begin{align}
    \frac{\dd \ln \mu_x}{\dd l}&=z-2\zeta+\frac{1}{16}g_\lambda^\perp \ ,\\
    \frac{\dd\ln \nu_\perp}{\dd l}&=z-4\ ,\\
    \frac{\dd \ln\lambda}{\dd l}&=\chi+z-\zeta-\frac{9}{8}g_\beta^\perp\ ,\\
    \frac{\dd \ln \beta}{\dd l}&=2\chi+z-\frac{9}{4}g_\beta^\perp\ ,\\
    \frac{\dd\ln D}{\dd l}&=\frac{z-\zeta-2\chi-d+1}{2}\ ,
\end{align}
where
\beq
    g_\lambda^\perp=\frac{D\lambda^2}{(\nu_\perp\mu_x)^{\frac{3}{2}}}\frac{S_{d-2}\Lambda^{d-7}}{(2\pi)^{d-1}} \ \ , \ \
    g_\beta^\perp=\frac{Db}{(\nu_\perp^3\mu_x)^{\frac{1}{2}}}\frac{S_{d-2}\Lambda^{d-7}}{(2\pi)^{d-1}} \ .
\eeq
In terms of these coupling coefficients, the RG flow equations are, 
\begin{align}
    \frac{\dd g_\lambda^\perp}{\dd l}&=\epsilon_T g_\lambda^\perp-\frac{3}{32}\left(g_\lambda^\perp \right)^2-\frac{9}{4}g_\lambda^\perp g_\beta^\perp \ ,\\
    \frac{\dd g_\beta^\perp}{\dd l}&=\epsilon_T g_\beta^\perp -\frac{9}{4}\left(g_\beta^\perp\right)^2 -\frac{1}{32}g_\lambda^\perp g_\beta^\perp
    \ ,
\end{align}
where $\epsilon_T = 7-d$ in this case. The resulting RG flow is depicted in Fig.~\ref{fig:RGflow}(b), showing that the multicritical behavior of this  TLP remains in the anisotropic equilibrium, albeit anisotropic LP universality class (red circle). However, upon further fine-tuning the $\beta$ term to zero, a new UC emerges (green square). The corresponding scaling exponents based on our 1-loop calculation are shown in Table~\ref{table}. 

As with the LPP, we can again compare our finding with the transverse critical transition of the KLS model when the $\beta$ term is absent \cite{janssen_ZPB86,leung_jstatphys86}. There, a new UC was also found, albeit with a different upper critical dimension, again due to the conservative nature of the noise term.

{\it LLP+TLP---}For completeness, we now consider the case of having both $\mu_x$ and $\mu_\perp$ fine tuned to zero, which at the bare level corresponds to the EOM:
\beq
\partial_t\phi+ \frac{\lambda}{2}\partial_x \phi^2 = -\nu_x\partial_x^4\phi - \nu_\bot\vnab_\bot^4\phi    
-\beta \phi^3 +f
 \ .
 \label{eq:LTeom}
\eeq
As detailed in the SM \cite{SM}, we find that the upper critical dimension is 10 and as in the LLP case, the RG flow from a 1-loop calculation indicates a divergent flow. Given the high upper critical dimension, it is of course difficult to draw any conclusion from such a perturbative treatment.

{\it Summary \& Outlook---}In summary, we analyzed the multicritical Lifshitz point (LP) behavior of the Active Malthusian Ising Model. We demonstrated that the system generically hosts two distinct LPs---longitudinal and transverse---and employed a dynamical renormalization-group (DRG) analysis within the $\epsilon$-expansion to characterize their universal behavior. At the longitudinal LP, activity drives a divergent RG flow, suggesting either a fluctuation-induced first-order transition or the existence of a strong-coupling fixed point. By contrast, the transverse LP remains in the equilibrium anisotropic LP universality class (UC). Remarkably, fine-tuning $\beta$ to zero reveals a new UC controlled solely by the active coupling $\lambda$. We further drew parallels to the critical behavior of the Katz–Lebowitz–Spohn model driven lattice gas. Given the broad and growing interest in LPs across physics, we expect our results to motivate future analytical, numerical, and experimental studies, deepening the understanding of multicritical behavior in active matter and beyond.

\vspace{.1in}
\begin{acknowledgments}
CFL thanks Patrick Jentsch for insightful discussions on Lifshitz points in active matter systems and for providing the software used to plot the RG flows in Fig.~2.
\end{acknowledgments}

%

\newpage\onecolumngrid

\noindent
{\bf Supplemental Material to: Universality class of Lifshitz points in Malthusian active Ising models }

\section{Longitudinal Lifshitz point}

In this section we drop the overcomplicated $\nu_x$, $\mu_\perp$ notations since we consider only the longitudinal case. We thus write $\nu$ and $\mu$ instead. 
\subsection{Linear Theory}
\subsubsection{Correlation function and linear scaling exponents}

Upon neglecting the two non-linear terms of the model (2), we can obtain the scaling behavior of the linear regime through the calculation of the correlation function of the field.
Let us first solve the linear equation in Fourier space :
\begin{equation}
    -i\omega\phi(\textbf{k},\omega)= -\mu k_\perp^2\phi(\textbf{k},\omega) - \nu k_x^4\phi(\textbf{k},\omega) + f(\textbf{k},\omega)
    \ ,
\end{equation}
where $k_x$ and $k_\perp$ correspond to momenta in the $x$ direction and in the transverse hyperplane. Thus,
\begin{equation}
\label{eq:EOMlin}
    \phi(\textbf{k},\omega) = G_0(\textbf{k},\omega)f(\textbf{k},\omega)
    \ ,
\end{equation}
with
\begin{equation}
\label{eq:G01}
    G_0(\textbf{k},\omega) = \frac{1}{\mu k_\perp^2 + \nu k_x^4 - i\omega}
    \ .
\end{equation}
We are interested in the correlation function of the field defined by $C_\phi(\textbf{r},t)=\langle\phi(\textbf{r},t)\phi(\textbf{0},0)\rangle$. We use the following notations:
\beq 
    \tilde{k}=(\textbf{k},\omega) \sep
        \int_{\tilde{k}} = \int_{-\infty}^{+\infty}\frac{d^dkd\omega}{(2\pi)^{d+1}} \ .
\eeq
Through an inverse Fourier transform we get 
    \begin{align}
        C_\phi(\textbf{r},t)&=\left\langle\int_{\tilde{k}}e^{i(\textbf{k}\cdot\textbf{r} - \omega t)} G_0(\textbf{k},\omega)\eta(\textbf{k},\omega)\int_{\tilde{k}'}G_0(\textbf{k'},\omega')\eta(\textbf{k'},\omega')\right\rangle \\
    &=\int_{\tilde{k}}\int_{\tilde{k}'}e^{i(\textbf{k}\cdot\textbf{r} - \omega t)} G_0(\textbf{k},\omega)G_0(\textbf{k'},\omega')2D\delta^d(\textbf{k} + \textbf{k'})\delta(\omega + \omega ') \\
    &=\int_{\tilde{k}}\frac{2De^{i(\textbf{k}\cdot\textbf{r} - \omega t)}}{(\mu k_\perp^2 + \nu k_x^4 - i\omega)(\mu k_\perp^2 + \nu k_x^4 + i\omega)}
\     .
    \end{align}    
We want to extract the $\mathbf{r_\perp}$ dependence of the expression, thus using the changes of variable :
\begin{equation}
    k_\perp = \frac{K_\perp}{\lvert \mathbf{r_\perp}\rvert} \sep k_x = \frac{K_x}{\lvert \mathbf{r_\perp}\rvert^{\frac{1}{2}}} \sep \omega = \frac{\Omega}{\lvert \mathbf{r_\perp}\rvert^2}
    \ .
\end{equation}
We finally get the expression:
\begin{align}
    C_\phi(\textbf{r},t)&=\lvert \mathbf{r_\perp}\rvert^{\frac{5}{2}-d}\int_{\tilde{K}}\frac{2De^{i(\mathbf{K_\perp}\cdot \mathbf{u}  + xK_x\lvert \mathbf{r_\perp}\rvert^{-\frac{1}{2}} - t\Omega\lvert \mathbf{r_\perp}\rvert^{-2})}}{(\mu K_\perp^2 + \nu K_x^4 - i\Omega)(\mu K_\perp^2 + \nu K_x^4 + i\Omega)}\nonumber \\
    &= \lvert \mathbf{r_\perp}\rvert^{\frac{5}{2}-d}S\left(\frac{x}{\lvert \mathbf{r_\perp}\rvert^{\frac{1}{2}}},\frac{t}{\lvert \mathbf{r_\perp}\rvert^2}\right) \ ,
\end{align}
where $S$ is a scaling function. This gives us the scaling exponents in the linear regime:
\begin{equation}
    \chi_L^{\rm lin}=\frac{5}{4}-\frac{d}{2} \sep \zeta_L^{\rm lin}=\frac{1}{2} \sep  z_L^{\rm lin}=2
    \ .
\end{equation}

\subsubsection{Upper critical dimensions}
We can now deduce the upper critical dimensions associated with the two non-linear terms of the model upon substituting the linear exponents into the full equation. To do so, we rescale the equation through the following relations:
\beq
    \mathbf{r_\perp} \longrightarrow \ee^\ell\mathbf{r_\perp} \sep x \longrightarrow \ee^{\zeta_L^{\rm lin}\ell} x \sep t \longrightarrow \ee^{z_L^{\rm lin}\ell} t \sep \phi \longrightarrow \ee^{\chi_L^{\rm lin}\ell} \phi 
    \ ,
\eeq
to obtain
\beq
    \partial_t\phi = \mu\nabla_\perp^2\phi - \nu\partial_x^4\phi + \frac{\lambda}{2}\ee^{(\chi - \zeta +z)\ell}\partial_x(\phi^2) - b\ee^{(2\chi + z)\ell}\phi^3 + f
    \ .
\eeq
By construction, the linear terms of the equation are invariant under rescaling, however, we see that:
\begin{itemize}
    \item The $\lambda$ term is rescaled by a factor $\ee^{(\chi-\zeta+z)\ell} = \ee^{(\frac{11}{4}-\frac{d}{2})\ell}$ which means that its upper critical dimension is $d=\frac{11}{2}=5.5$. This term flows to zero above this critical dimension and to infinity below it.
    \item The $b$ term is rescaled by a factor $\ee^{(2\chi + z)\ell}=\ee^{(\frac{9}{2}-d)\ell}$ which means similarly that its upper critical dimension is $d = 4.5$.
\end{itemize}

\subsection{Dynamical Renormalization Group analysis in dimension $5.5 - \epsilon$}

\subsubsection{Non-linear equation in Fourier space}
We will now consider the model in dimension $5.5 - \epsilon$, neglecting the $b$ term, as we have shown it to be irrelevant when $\epsilon$ is small. We write in Fourier space:
\beq
    -i\omega\phi(\textbf{k},\omega)= -\mu k_\perp^2\phi(\textbf{k},\omega) - \nu k_x^4\phi(\textbf{k},\omega) + i\frac{\lambda}{2}k_x\int_{\tilde{q}}\phi(\tilde{q})\phi(\tilde{k}-\tilde{q}) + f(\textbf{k},\omega)
    \ .
\eeq
Using the definition of $G_0(\tilde{k})$ (\ref{eq:G01}):
\begin{equation}
\label{eq:EOMLongi}
    \phi(\tilde{k}) = G_0(\tilde{k})f(\tilde{k}) + i\frac{\lambda}{2}k_xG_0(\tilde{k})\int_{\tilde{q}}\phi(\tilde{q})\phi(\tilde{k}-\tilde{q})
    \ ,
\end{equation}
which can be written through the general propagator $G(\tilde{k})$ :
\begin{equation}
    \phi(\tilde{k}) = G(\tilde{k})f(\tilde{k}) \ .
\end{equation}

This recursive relation can be approximated to whichever order of $\lambda$. To get corrections to the coefficients $\mu$, $\nu$, $\lambda$ and $D$, we first map this equation into diagrams and then calculate different quantities to second order in $\lambda$ (corresponding to one-loop expansions in the diagrams).

\subsubsection{Diagrammatic expansions}
In order to map the equation to diagrams, we use the following conventions :
\begin{equation}
    G(\tilde{k)} = \raisebox{-0.5cm}{\includegraphics[scale=0.2]{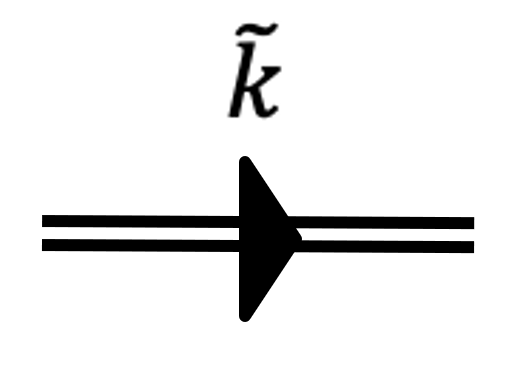}}
\end{equation}
\begin{equation}
    G_0(\tilde{k)} = \raisebox{-0.45cm}{\includegraphics[scale=0.15]{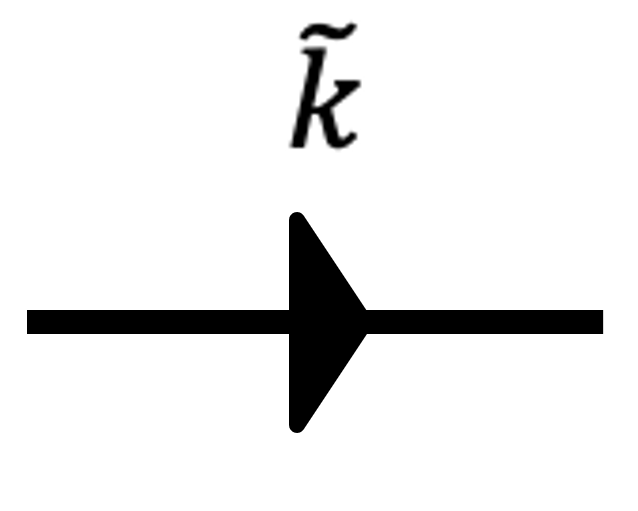}}
\end{equation}
\begin{equation}
    2D = \raisebox{-0.5cm}{\includegraphics[scale=0.18]{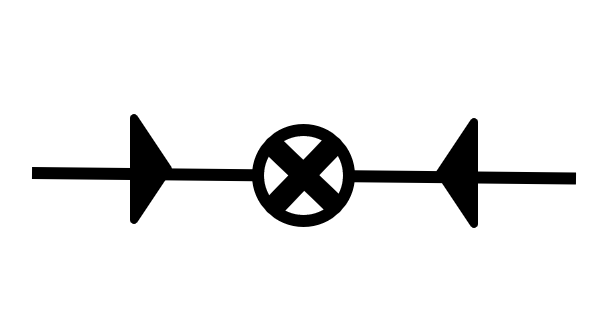}}
\end{equation}
\begin{equation}
    i\frac{\lambda}{2}k_x\int_{\tilde{q}} = \raisebox{-0.8cm}{\includegraphics[scale=0.12]{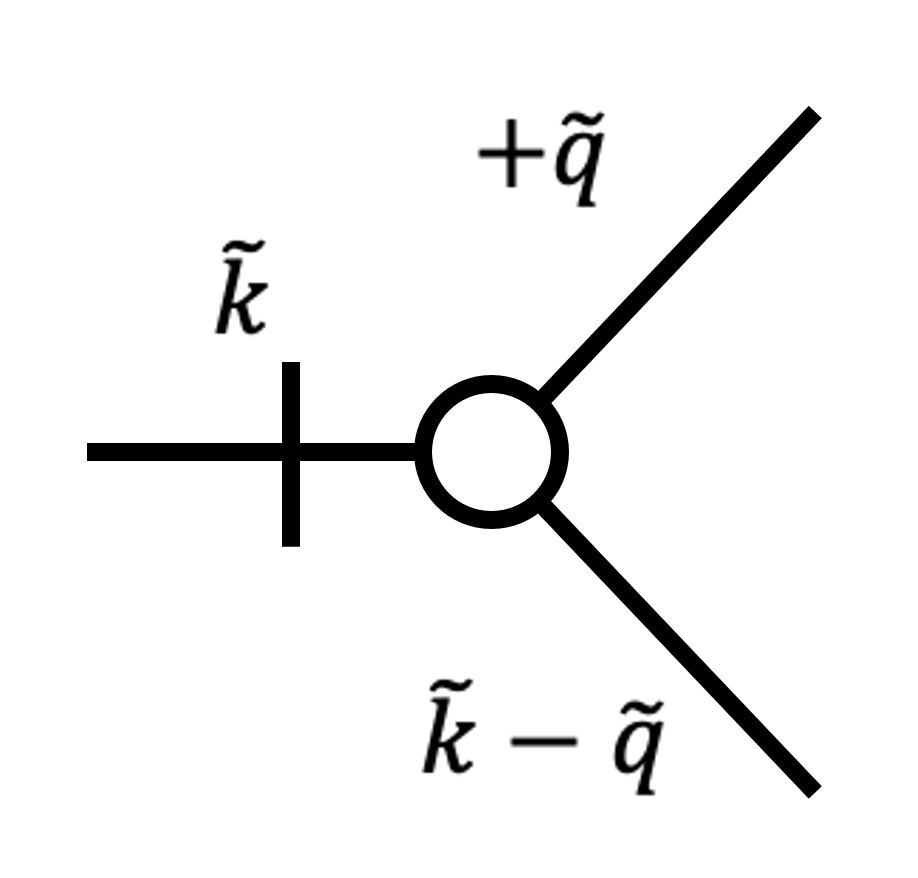}}
\end{equation}
We now deduce through diagram calculation the three following one-loop approximations for the propagator, vertex and noise :
\begin{equation}
\label{eq:prop1}
    \raisebox{-0.5cm}{\includegraphics[scale=0.2]{Diagrams/Propagator_k.PNG}} = \raisebox{-0.45cm}{\includegraphics[scale=0.15]{Diagrams/Propagator_linear_k.PNG}} + 4\raisebox{-0.82cm}{\includegraphics[scale=0.2]{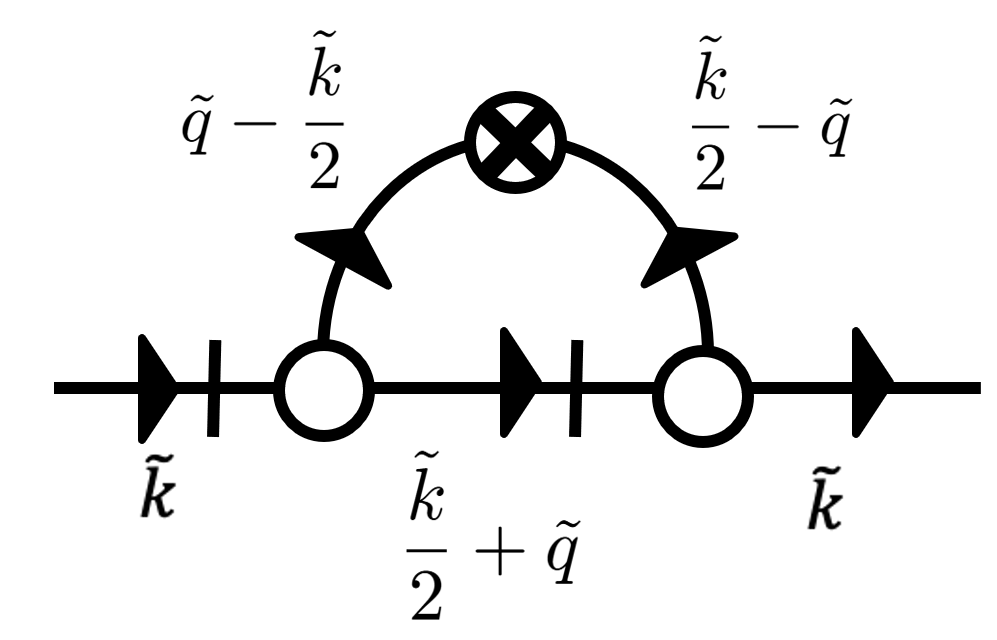}}
\end{equation}
\begin{align}
\label{eq:vertexdiagram1}
    \raisebox{-0.85cm}{\includegraphics[scale=0.2]{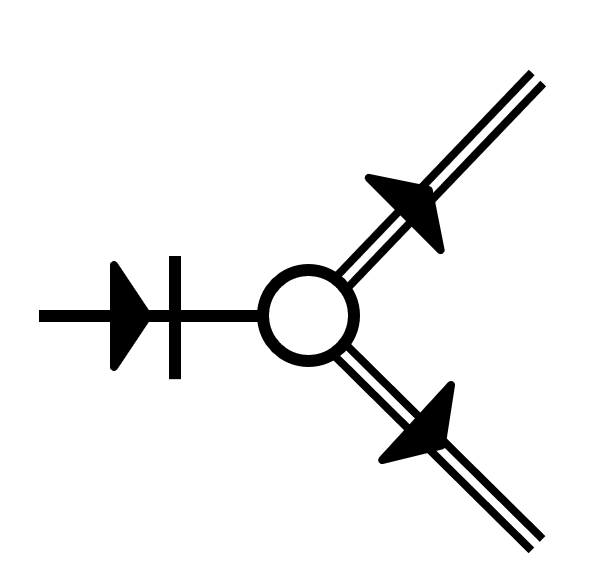}} = &\raisebox{-0.95cm}{\includegraphics[scale=0.2]{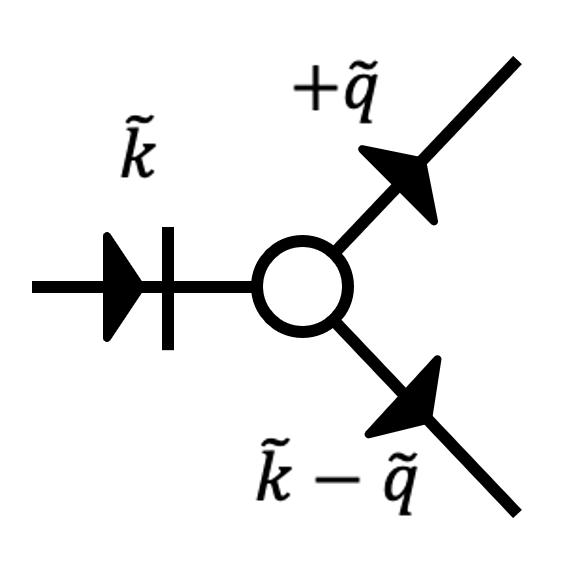}} + 4 \raisebox{-1.25cm}{\includegraphics[scale=0.2]{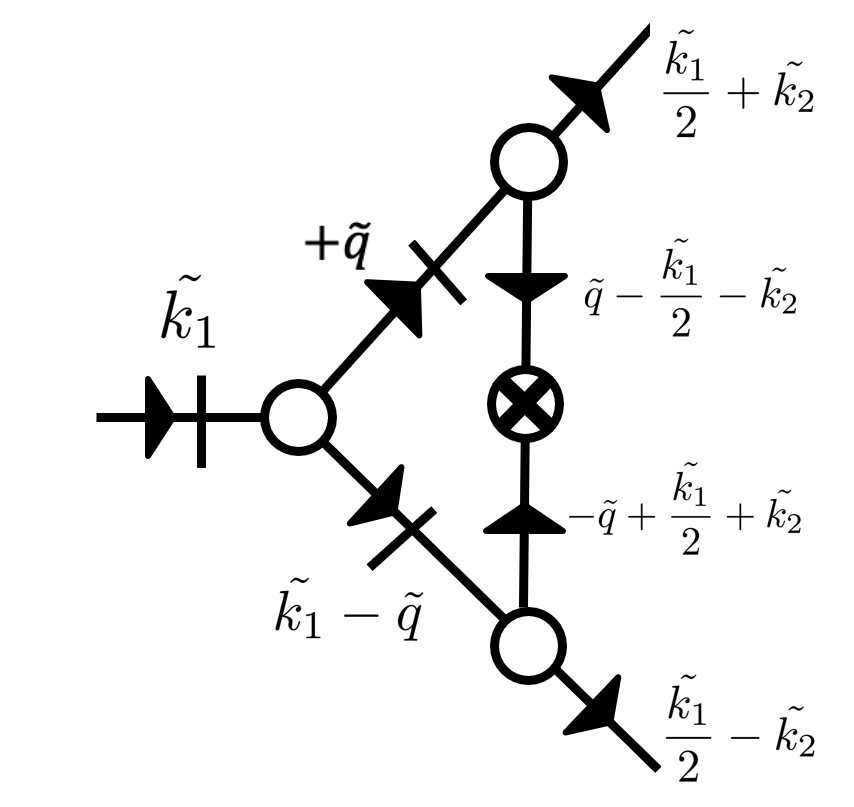}} + 4 \raisebox{-1.25cm}{\includegraphics[scale=0.2]{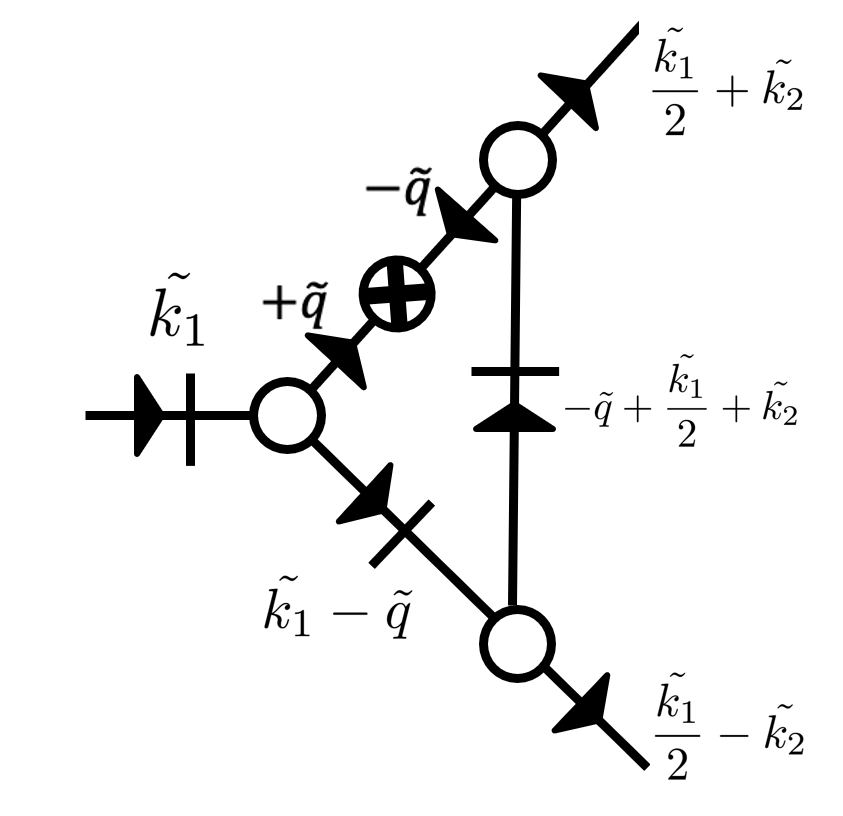}}\nonumber\\
    &+ 4 \raisebox{-1.5cm}{\includegraphics[scale=0.2]{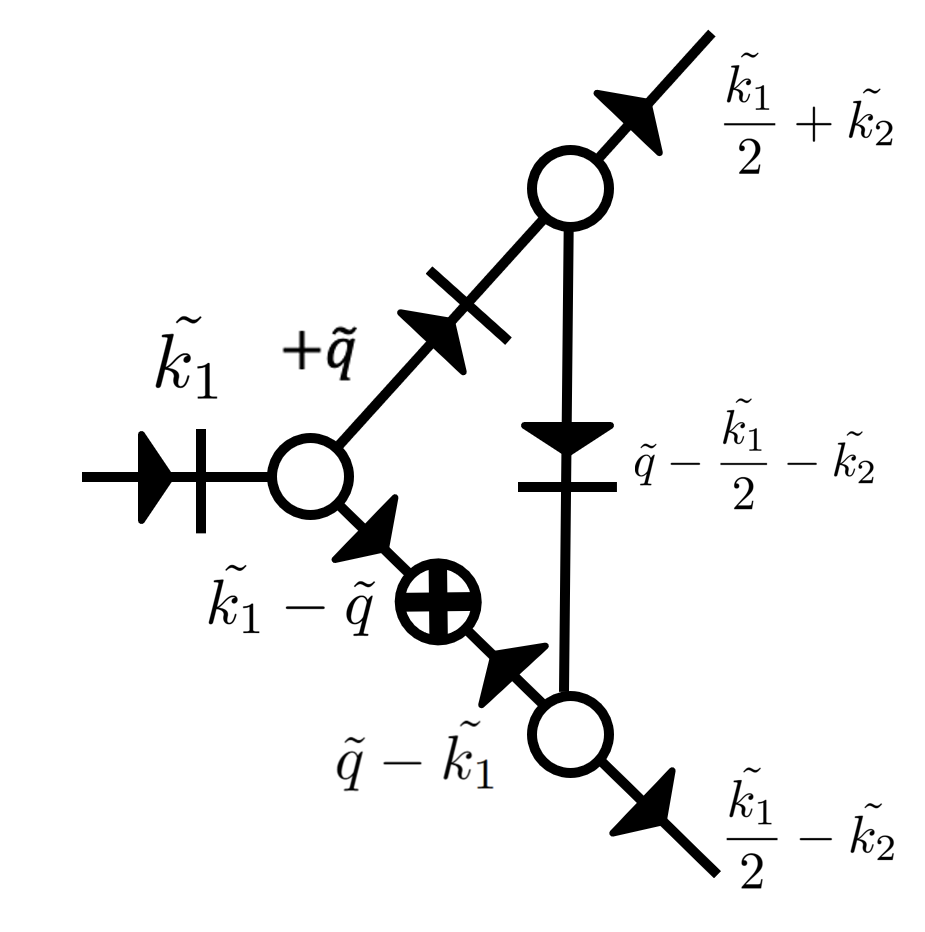}}
\end{align}
\begin{equation}
\label{eq:noisediagram}
    \raisebox{-0.4cm}{\includegraphics[scale=0.18]{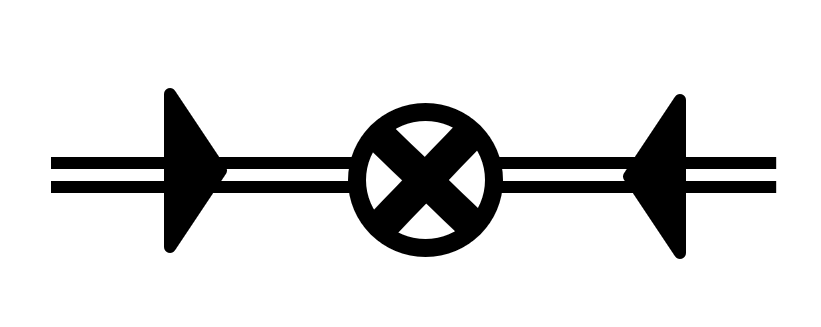}} = \raisebox{-0.6cm}{\includegraphics[scale=0.25]{Diagrams/Noise.PNG}} + 2\raisebox{-1cm}{\includegraphics[scale=0.18]
    {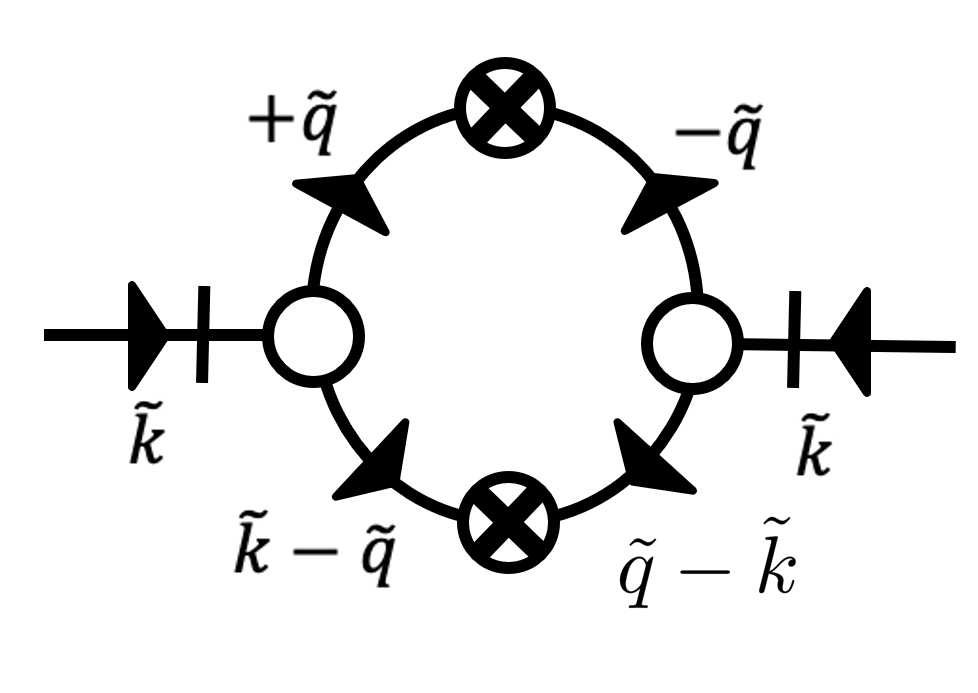}}
\end{equation}

\subsubsection{Expansion of the propagator}
At 1-loop, the diagrammatic equation (\ref{eq:prop1}) in Fourier space can be written as
\begin{equation}
\label{eq:prop2}
    G\left(\tilde{k} \right) = G_0(\tilde{k}) - 2D\lambda^2G_0^2\left(\tilde{k}\right)k_x\int_{\tilde{q}}\left(\frac{k_x}{2}+q_x\right)G_0\left(\frac{\tilde{k}}{2}+\tilde{q}\right)G_0\left(\frac{\tilde{k}}{2}-\tilde{q}\right)G_0\left(\tilde{q}-\frac{\tilde{k}}{2}\right) \ ,
\end{equation}
where 
\begin{equation}
        \int_{\tilde{q}} = \int_{\mathbf{q_\perp}}^\Lambda \int_{q_x} \int_\Omega
        =\int_{|\mathbf{q_\perp|}\in[\Lambda e^{-dl}, \Lambda]} \frac{d^{d-1}\mathbf{q_\perp}}{(2\pi)^{d-1}}\int_{-\infty}^{+\infty}\frac{dq_x}{2\pi} \int_{-\infty}^{+\infty}\frac{d\Omega}{2\pi} \ .
\end{equation}
In the above,
 the integral  over  $\mathbf{q_\perp}$ is  restrained to the momentum shell defined by $\lvert \bq_\perp\rvert \in [\Lambda e^{-dl},\Lambda]$. Here the  upper bound, $\Lambda$, denotes the Wilsonian RG cutoff and is the central quantity of the theory. It is the equivalent of a lattice size for a discrete model, implying that the model must have an upper bound in momentum. According to the usual process of renormalization we only integrate out a momentum shell, given that the integral diverges when the lower bound is zero. The introduction of the ``RG time" $l$ in the lower bound of the integral is the source of the flow equations that are uncovered below.

Due to the factor  $k_x$ (indicated by the vertical bar) in front of the 1-loop diagram, the graphical correction cannot contribute to $\mu$. 

The correction for $\nu$ is more complicated as we need to expand the integrand in (\ref{eq:prop2}) to third order in $k_x$. To do so we first expand $G_0$  to third order in $k_x$. For instance,
\begin{align}
\nonumber
    G_0\left(\frac{\tilde{k}}{2}+\tilde{q}\right) = G_0(&\tilde{q})-2\nu k_xq_x^3G_
    0^2(\tilde{q}) +\left[-\frac{3}{2}\nu k_x^2q_x^2+4\nu^2k_x^2q_x^6G_0(\tilde{q})\right]G_0^2(\tilde{q}) 
    \\ &+\left[-\frac{1}{2}\nu k_x^3q_x+6\nu^2k_x^3q_x^5G_0(\tilde{q})-8\nu^3k_x^3q_x^9G_0^2(\tilde{q})\right]G_0^2(\tilde{q}) +O(k_x^4) \ , \label{eq:linpropexp}
    \\    
    \nonumber
    G_0\left(\frac{\tilde{k}}{2}-\tilde{q}\right) = G_0(-&\tilde{q})+2\nu k_xq_x^3G_
    0^2(-\tilde{q}) +\left[-\frac{3}{2}\nu k_x^2q_x^2+4\nu^2k_x^2q_x^6G_0(-\tilde{q})\right]G_0^2(-\tilde{q}) \\ &+\left[\frac{1}{2}\nu k_x^3q_x-6\nu^2k_x^3q_x^5G_0(-\tilde{q})+8\nu^3k_x^3q_x^9G_0^2(-\tilde{q})\right]G_0^2(-\tilde{q}) +O(k_x^4) \ .
\end{align}

We now write the integral as an expansion of $k_x$:
\beq
\int_{\tilde{q}}\left(\frac{k_x}{2}+q_x\right)G_0\left(\frac{\tilde{k}}{2}+\tilde{q}\right)G_0\left(\frac{\tilde{k}}{2}-\tilde{q}\right)G_0\left(\tilde{q}-\frac{\tilde{k}}{2}\right) =
A_0+A_1k_x+A_2 k_x^2+A_3k_x^3+\cO(k_x^4)\ ,
\eeq
where the $A$'s are independent of $k_x$.  Due to the oddness of $q_x$ in the integrals denoted by $A_0$ and $A_3$, they are both zero. $A_2$ is nonzero, but since it is the term that we fine tune to zero to get to the LLP, we will ignore it here. Finally, we need to calculate $A_3$. To do so, we first integrate out $\Omega$, with the help of the following formulas (established with the residue theorem):
\begin{align}
    &\int_{-\infty}^\infty \frac{dz}{2\pi} \frac{1}{(a-iz)^\gamma(a+iz)}=\frac{1}{(2a)^\gamma} \ ,\label{eq:res1}\\
    &\int_{-\infty}^\infty \frac{dz}{2\pi} \frac{1}{(a-iz)^\gamma(a+iz)^2}=\frac{\gamma}{(2a)^{\gamma+1}}\ , \label{eq:res2}\\
    &\int_{-\infty}^\infty \frac{dz}{2\pi} \frac{1}{(a-iz)^\gamma(a+iz)^3}=\frac{\gamma(\gamma+1)}{2^{\gamma+3}a^{\gamma+2}}\ , \label{eq:res3}\\
    &\int_{-\infty}^\infty \frac{dz}{2\pi} \frac{1}{(a-iz)^\gamma(a+iz)^4}=\frac{\gamma(\gamma+1)(\gamma+2)}{3\times 2^{\gamma+4}a^{\gamma+3}} \ , \label{eq:res4}
\end{align}
yielding:
\begin{equation}
  A_3  = \int_q\left( -\frac{\mu q_x^2}{4\Gamma^3}-\frac{7\mu^2q_x^6}{4\Gamma^4}+\frac{2\mu^3q_x^{10}}{\Gamma^5} \right)\ ,
\end{equation}
where 
\begin{equation}
    \Gamma = \mu q_\perp^2 + \nu q_x^4
     \ .
\end{equation}

After integrating by parts the above is reduced to :
\begin{equation}
   A_3 = -\frac{15}{32}\nu k_x^3\int_q \frac{q_x^2}{(\mu q_\perp^2+\nu q_x^4)^3} \ ,
\end{equation}
which can evaluated as follows:
\begin{equation}
    -\frac{15}{32}k_x^3\int_{\mathbf{q_\perp}}^\Lambda \frac{1}{q_\perp^{\frac{9}{2}}}\left(\frac{\nu}{\mu^9}\right)^{\frac{1}{4}}\int_{-\infty}^{+\infty}\frac{du}{(2\pi)}\frac{u^2}{(1+u^4)^3} = -\frac{75}{2048\sqrt{2}}k_x^3\left(\frac{\nu}{\mu^9}\right)^{\frac{1}{4}}\int_{\mathbf{q_\perp}}^\Lambda \frac{1}{q_\perp^{\frac{9}{2}}} \ .
\end{equation}

Injecting this result into (\ref{eq:prop2}), we have
\begin{equation}
    G^<(k_\perp=0,k_x,\omega=0) = G_0(k_x) + \frac{75}{1024\sqrt{2}}D\lambda^2\left(\frac{\nu}{\mu^9}\right)^\frac{1}{4}G_0^2(k_x)k_x^4\int_{q_\perp}^\Lambda\frac{1}{q_\perp^\frac{9}{2}}
    \ .
\end{equation}
The $\mathbf{q_\perp}$ integral can now be written as
\beq
    \int_{|\mathbf{q_\perp|}\in[\Lambda e^{-dl}, \Lambda]} \frac{d\mathbf{q_\perp}}{(2\pi)^{d-1}}\frac{1}{q_\perp^\frac{9}{2}} = \frac{S_{d-1}}{(2\pi)^{d-1}}\int_{\Lambda e^{-dl}}^\Lambda \frac{dq_x}{q_x^{\frac{9}{2}}}=\frac{S_{d-1}\Lambda^{d-\frac{11}{2}}}{(2\pi)^{d-1}} \ .
\eeq
Where we introduced $S_d$ the area of a hypersphere of dimension $d$.

We finally get the correction:
\begin{equation}
\label{eq:Gnu}
    G^<(k_\perp=0,k_x,\omega=0) = G_0(k_x) + \frac{75}{1024\sqrt{2}}D\lambda^2\left(\frac{\nu}{\mu^9}\right)^\frac{1}{4}G_0^2(k_x)k_x^4\frac{S_{d-1}\Lambda^{d-\frac{11}{2}}}{(2\pi)^{d-1}}
    \ .
\end{equation}

In order to obtain a correction for $\nu$ we write $ G^<(k_\perp=0,k_x,\omega=0) = \frac{1}{\nu^{<}k_x^4}$ by identification of the form of the $G_0$ term, where $\nu^<$ indicates the $\nu$ coefficient modified by taking into account the corrections for $|\mathbf{q_\perp}|\in [\Lambda e^{-dl},\Lambda]$. Using (\ref{eq:Gnu}) we thus have after a simple series expansion:
\begin{equation}
\label{eq:NuSupexp}
    \nu^<=\nu\left[1 -\frac{75}{1024\sqrt{2}}\frac{S_{d-1}}{(2\pi)^{d-1}}\frac{D\lambda^2}{\left(\nu^3\mu^9\right)^\frac{1}{4}}\Lambda^{d-5.5}d\ell\right]
    \ .
\end{equation}

\subsubsection{Expansion of the vertex}
The diagrammatic expansion of the vertex (\ref{eq:vertexdiagram1}) yields
\beq
    \lambda^<=\lambda(1+\Gamma_a + \Gamma_b + \Gamma_c)
    \ .
\eeq
With the following expressions :
\begin{align}
    &\Gamma_a = -2D\lambda^2\int_{\tilde{q}} q_x(k_{1x}-q_x)G_0(\tilde{q})G_0(\tilde{k_1}-\tilde{q})G_0(\tilde{q}-\frac{\tilde{k_1}}{2}-\tilde{k_2})G_0(-\tilde{q}+\frac{\tilde{k_1}}{2}+\tilde{k_2}) \ , \label{eq:gammaA}
    \\
    &\Gamma_b = -2D\lambda^2\int_{\tilde{q}} (k_{1x}-q_x)(\frac{\tilde{k_1}}{2}+\tilde{k_2}-\tilde{q})G_0(\tilde{q})G_0(-\tilde{q})G_0(\tilde{k_1}-\tilde{q})G_0(\frac{\tilde{k_1}}{2}+\tilde{k_2}-\tilde{q})
    \ , \label{eq:gammaB}
     \\
    &\Gamma_c = -2D\lambda^2\int_{\tilde{q}} q_x(q_x-\frac{k_{1x}}{2}-k_{2x})G_0(\tilde{q})G_0(\tilde{q}-\tilde{k_1})G_0(\tilde{k_1}-\tilde{q})G_0(\tilde{q} - \frac{\tilde{k_1}}{2}-\tilde{k_2})
    \ . \label{eq:gammaC}
\end{align}
We demonstrate that the sum of the three is null. Given that we only need the zeroth order correction in $k_x$ (any other term goes to zero in the hydrodynamic limit) we have
\begin{align}
    \Gamma_a &=-2D\lambda^2\int_{\tilde{q}} q_x(k_{1x}-q_x)G_0(\tilde{q})G_0(\tilde{k_1}-\tilde{q})G_0(\tilde{q}-\frac{\tilde{k_1}}{2}-\tilde{k_2})G_0(-\tilde{q}+\frac{\tilde{k_1}}{2}+\tilde{k_2}) \\
    &=2D\lambda^2\int_{\tilde{q}}q_x^2G_0^2(\tilde{q})G_0^2(-\tilde{q}) +O(k_x)
    \ .
\end{align}
We integrate over $\Omega$ using a simple contour integral, yielding
\beq
    \Gamma_a =2D\lambda^2\int_{q} \frac{q_x^2}{4(\mu q_\perp^2+\nu q_x^4)^3} +O(k_x)
     \ .
\eeq
The exact same calculation applies to $\Gamma_b$ and $\Gamma_c$ and yields :
\beq
    \Gamma_b = \Gamma_c =-2D\lambda^2\int_q \frac{q_x^2}{8(\mu q_\perp^2+\nu q_x^4)^3} +O(k_x)
    \ ,
\eeq
which thus yields the result
\beq
    \Gamma_a + \Gamma_b + \Gamma_c = 0
    \ .
\eeq

The three terms thus exactly compensate, leading to no correction for the $\lambda$ coefficient after the one-loop calculation:
\begin{equation}
    \lambda^<=\lambda
    \ .
\end{equation}

\subsubsection{Expansion of the noise}
The expansion of the noise gives:
\begin{equation}
\label{eq:noisesupexp}
    D^<=D +D^2\lambda^2k_x^2\int_{\tilde{q}} G_0(\tilde{q})G_0(-\tilde{q})G_0(\tilde{k}-\tilde{q})G_0(\tilde{q}-\tilde{k})
\end{equation}
The right term cannot be renormalized because it is a second order term in $k_x$, which always goes to zero in the hydrodynamic limit. Thus we yield no correction for the noise:
\begin{equation}
    D^<=D
    \ .
\end{equation}

\subsubsection{Flow equations and divergence of the flow}
The corrections we got correspond to the rescaling of the system by a coefficient $e^{dl}$. Thus the renormalized coefficients are related to the rescaled coefficients through rescaling. We get for example:
\beq
    \tilde{\mu}=e^{dl(z-2)}\mu^<
    \ .
\eeq
Where the exponents are the same as in power counting analysis we made earlier. Thus we obtain the flow equation :
\begin{equation}
\label{eq:flowLongiMu}
    \frac{d\mu}{dl}=\mu(z-2)
    \ .
\end{equation}
Similarly we obtain the other flow equations for all the coefficients:
\beqn
\label{eq:flowLongiNu}
    \frac{d\nu}{dl}&=&\nu\left(z-4\zeta -\frac{75}{1024\sqrt{2}}\frac{D\lambda^2}{(\nu^3\mu^9)^\frac{1}{4}}\frac{S_{d-2}\Lambda^{d-5.5}}{(2\pi)^{d-1}}\right)
    \ ,
    \\
    \frac{d\lambda}{dl}&=&\lambda(\alpha+z-\zeta) \label{eq:flowLongiLambda}
    \ ,
    \\
        \frac{dD}{dl}&=&\frac{D}{2}(z-\zeta-2\alpha-d+1)\ . \label{eq:flowLongiD}
\eeqn
We are interested in the fixed points of the renormalization group, thus of the system of flow equations \ref{eq:flowLongiMu}-\ref{eq:flowLongiD}. We introduce the coupling constant $g_\lambda^x$, which is involved in equation \ref{eq:flowLongiNu}:
\begin{equation}
\label{eq:couplingLongi}
    g_\lambda^x=\frac{D\lambda^2}{(\nu^3\mu^9)^\frac{1}{4}}\frac{S_{d-2}\Lambda^{d-5.5}}{(2\pi)^{d-1}} \ .
\end{equation}

The flow equations fixed points for $\mu$ and $D$ yield the following scaling results:
\begin{equation}
    z=2 \ ,
\end{equation}
\begin{equation}
    2\alpha+\zeta=3-d \ .
\end{equation}
By applying a $log$ function to \ref{eq:couplingLongi} and differentiating, one gets:
\begin{equation}
    \frac{1}{g_\lambda^x}\frac{dg_\lambda^x}{dl} = \frac{1}{D}\frac{dD}{dl} +\frac{2}{\lambda}\frac{d\lambda}{dl}-\frac{3}{4\nu}\frac{d\nu}{dl}-\frac{9}{4\mu}\frac{d\mu}{dl} \ .
\end{equation}
By injecting \ref{eq:flowLongiMu}-\ref{eq:flowLongiD} one can finally derive the flow equations for the coupling constant:
\begin{equation}
\label{eq:couplingFlowLongi}
    \frac{dg_\lambda^x}{dl}= \epsilon g_\lambda^x  + \frac{225}{4096\sqrt{2}}(g_\lambda^x)^2 \ .
\end{equation}
Where $\epsilon=5.5-d$. This equation leads to two mathematical fixed points:
\begin{align}
    g_{\lambda,1}^{x,*}&=0 \ , \\
    g_{\lambda,2}^{x,*}&=-\frac{4096\sqrt{2}}{225}\epsilon \ .
\end{align}
Given the definition of $g_\lambda^x$, it should only take positive values. However, we notice that $g_{\lambda}^{x,*}$ is either null or negative, leading to the conclusion that the only physical fixed point is the linear fixed point $g_{\lambda}^{x,*}=0$. Given equation \ref{eq:couplingFlowLongi} we also notice that the coupling constant diverges to infinity through the RG flow from any strictly positive value, which is interpreted as a discontinuous phase transition (first type).

\subsection{Uncontrolled 1-loop DRG analysis for $d<4.5$}
Our perturbative method is valid only in the small $\epsilon$ regime. To probe the impact of the $\beta$, we repeat our 1-loop calculation  for $d<4.5$ and with the $\beta$-vertex included. Given the large $\epsilon>1$, this approach is out of the perturbative regime is thus uncontrolled. However, it does enable to gauge qualitatively into the potential impact of the $\beta$ term.

\subsubsection{Diagrammatic expansion and flow equations}
We use the diagrammatic representation:
\begin{equation}
    -\beta \int_{\tilde{p}_1}\int_{\tilde{p}_2} = \raisebox{-0.65cm}{\includegraphics[scale=0.22]
    {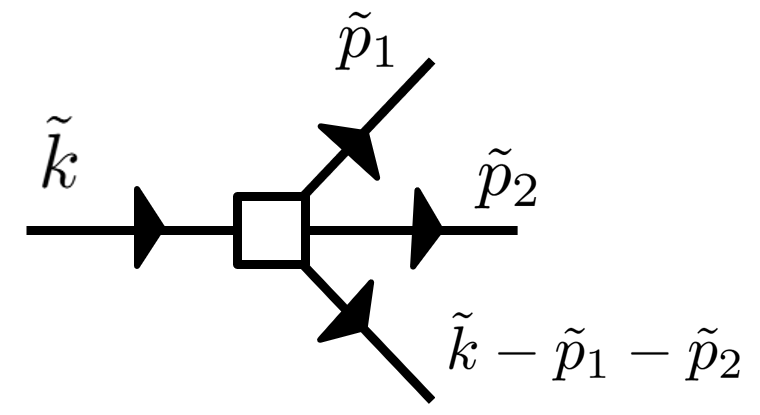}} \ .
\end{equation}

Comparing to the $5.5-\epsilon$ dimension analysis, corrections to the propagator \ref{eq:prop1} and noise \ref{eq:noisediagram} are unchanged by the addition of a $\beta$ term. However the 3-branch vertex correction \ref{eq:vertexdiagram1} is modified, and a 4-branch vertex has to be analyzed. We thus consider the following equations :
\begin{align}
\label{eq:vertex3diagramUncontrolled}
    \raisebox{-0.85cm}{\includegraphics[scale=0.2]{Diagrams/Vertex.PNG}} = &\raisebox{-0.95cm}{\includegraphics[scale=0.2]{Diagrams/Vertex_linear_k.PNG}} + 4 \raisebox{-1.25cm}{\includegraphics[scale=0.2]{Diagrams/gamma_A.PNG}} + 4 \raisebox{-1.25cm}{\includegraphics[scale=0.2]{Diagrams/gamma_B.PNG}}\nonumber\\
    &+ 4 \raisebox{-1.5cm}{\includegraphics[scale=0.2]{Diagrams/gamma_C.PNG}}+ 6 \raisebox{-0.95cm}{\includegraphics[scale=0.4]{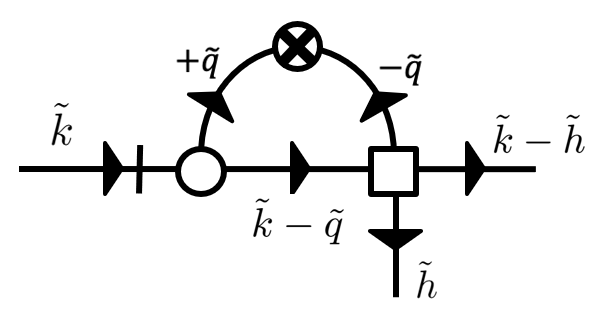}}\nonumber\\
    &+ 4 \raisebox{-0.95cm}{\includegraphics[scale=0.4]{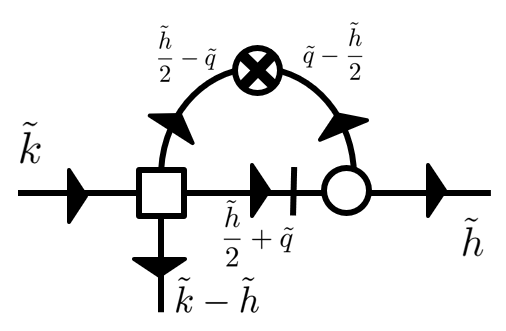}} \ ,
\end{align}

\begin{align}
\label{eq:vertex4diagramUncontrolled}
    \raisebox{-0.8cm}{\includegraphics[scale=0.4]{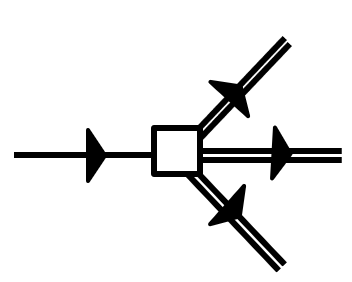}} = &\raisebox{-0.75cm}{\includegraphics[scale=0.25]{Diagrams/3_legs_basic.PNG}} + 18 \raisebox{-0.85cm}{\includegraphics[scale=0.45]{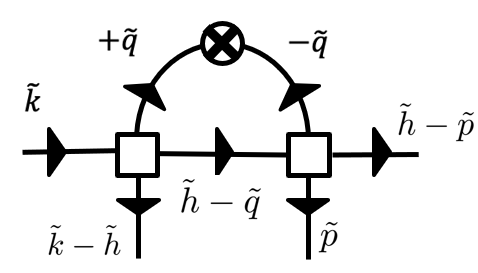}}\nonumber\\
    &+ 12 \raisebox{-1.45cm}{\includegraphics[scale=0.4]{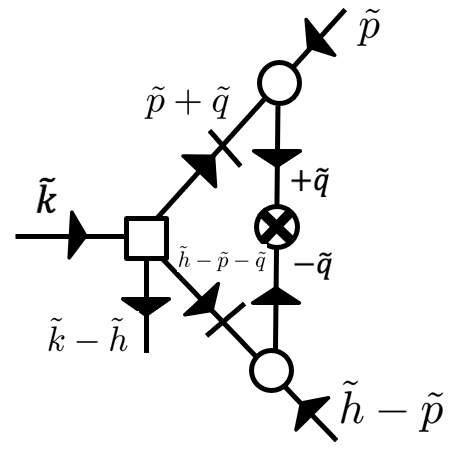}}+ 24 \raisebox{-1.6cm}{\includegraphics[scale=0.45]{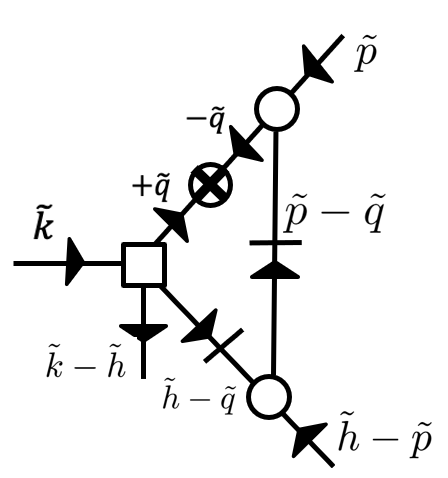}} \ .
\end{align}

\subsubsection{$\lambda$-vertex expansion and flow equation}
The four first terms in the vertex expansion are similar to the $d=5.5-\epsilon$ case, and yield no correction. We now analyze the two last, which we call respectively A- and B-graph.

The A-graph yields an overall correction to the equation:
\begin{align}
    \delta(\partial_t\phi)&=-6iD\lambda\beta G_0(\tilde{k})k_x\int_{\tilde{h}}G_0(\tilde{h})G_0(\tilde{k}-\tilde{h}) \int_{\tilde{q}} G_0(\tilde{q})G_0(-\tilde{q})G_0(\tilde{k}-\tilde{q}) \ ,
\end{align}
which is then compared to the linear term $i\frac{\lambda}{2}k_xG_0(\tilde{k})\int_{\tilde{h}}G_0(\tilde{h})G_0(\tilde{k}-\tilde{h})$. Thus this graph yields a correction to the $\lambda$ coefficient :
\begin{align}
    \delta\lambda_A &= -12D\lambda\beta\int_{\tilde{q}} G_0(\tilde{q})G_0^2(-\tilde{q}) \ ,
\end{align}
because we are only interested in the zeroth order in $\tilde{k}$ in the internal integral. Using previously introduced techniques, we yield:
\begin{align}
    \delta\lambda_A&=-\frac{9}{8\sqrt{2}}\frac{\beta D\lambda}{(\mu^7\nu)^{\frac{1}{4}}}\int_{\mathbf{q_\perp}}^{\Lambda} \frac{1}{q_\perp^{\frac{7}{2}}} \\
    &=-\frac{9}{8\sqrt{2}}\frac{\beta D\lambda}{(\mu^7\nu)^{\frac{1}{4}}}\frac{S_{d-2}\Lambda^{d-4.5}}{(2\pi)^{d-1}}dl \ .
\end{align}
The B-graph leads to an overall correction:
\begin{align}
    \delta(\partial_t\phi)&=-12iD\lambda\beta G_0(\tilde{k})\int_{\tilde{h}}G_0(\tilde{h})G_0(\tilde{k}-\tilde{h}) \int_{\tilde{q}} (h_x-q_x)G_0(\tilde{q})G_0(-\tilde{q})G_0(\tilde{k}-\tilde{q})
    \ ,
\end{align}
which has again to be compared to the linear term. To do so, we first note that
\begin{equation}
    \int_{\tilde{h}} h_xG_0(\tilde{h})G_0(\tilde{k}-\tilde{h}) = \int_{\tilde{h}} (k_x-h_x)G_0(\tilde{h})G_0(\tilde{k}-\tilde{h})
\end{equation}
by simple change of variable, and thus:
\begin{equation}
    \int_{\tilde{h}} h_xG_0(\tilde{h})G_0(\tilde{k}-\tilde{h}) = \frac{1}{2}k_x\int_{\tilde{h}} G_0(\tilde{h})G_0(\tilde{k}-\tilde{h})
    \ .
\end{equation}
After some more usual calculation this relation allows us to obtain:
\begin{equation}
    \delta\lambda_B=\delta\lambda_A +\frac{9}{32\sqrt{2}}\frac{\beta D\lambda}{(\mu^7\nu)^{\frac{1}{4}}}\frac{S_{d-2}\Lambda^{d-4.5}}{(2\pi)^{d-1}}dl
     \ .
\end{equation}
Putting both corrections together, we finally yield:
\begin{equation}
    \lambda^<=\lambda\left( 1-\frac{63}{32\sqrt{2}}\frac{\beta D}{(\mu^7\nu)^{\frac{1}{4}}}\frac{S_{d-2}\Lambda^{d-4.5}}{(2\pi)^{d-1}}dl \right)
     \ .
\end{equation}
We finally yield (after using the rescaling relation):
\begin{equation}
    \frac{d\lambda}{dl}=\lambda\left(\alpha+z-\zeta-\frac{63}{32\sqrt{2}}\frac{\beta D}{(\mu^7\nu)^{\frac{1}{4}}}\frac{S_{d-2}\Lambda^{d-4.5}}{(2\pi)^{d-1}}\right)
    \ .
\end{equation}

\subsubsection{$\beta$-vertex expansion and flow equation}
The zeroth order term of the expansion is simply written $-\beta\int_{\tilde{h},\tilde{p}}G_0(\tilde{p})G_0(\tilde{h}-\tilde{p})G_0(\tilde{k}-\tilde{h})$, and the correction terms are shown graphically in (\ref{eq:vertex4diagramUncontrolled}). We name the three 1-loop correction terms A-, B- and C-graphs.

The A-graph gives a correction to the equation:
\begin{equation}
    \delta(\partial_t\phi)=36\beta^2DG_0(\tilde{k})\int_{\tilde{h},\tilde{p}}G_0(\tilde{k}-\tilde{h})G_0(\tilde{p})G_0(\tilde{h}-\tilde{p}) \int_{\tilde{q}} G_0(\tilde{q})G_0(-\tilde{q})G_0(\tilde{h}-\tilde{q})
    \ .
\end{equation}
At zeroth order in $\tilde{h}$ we get:
\begin{equation}
    \delta\beta_A=-36\beta^2D\int_{\tilde{q}}G_0(\tilde{q})G_0^2(-\tilde{q})
    \ ,
\end{equation}
which thus yields after integration:
\begin{equation}
    \delta\beta_A=-\frac{27}{8\sqrt{2}}\frac{\beta^2D}{(\mu^7\nu)^{\frac{1}{4}}}\frac{S_{d-2}\Lambda^{d-4.5}}{(2\pi)^{d-1}}dl
    \ .
\end{equation}

We will now show that the B- and C-graph cancel each other. The B-graph correction is:
\begin{equation}
    \delta(\partial_t\phi)=6\beta D\lambda^2G_0(\tilde{k})\int_{\tilde{h},\tilde{p}}G_0(\tilde{k}-\tilde{h})G_0(\tilde{p})G_0(\tilde{h}-\tilde{p}) \int_{\tilde{q}} (p_x+q_x)(h_x-p_x-q_x)G_0(\tilde{q})G_0(-\tilde{q})G_0(\tilde{p}+\tilde{q})G_0(\tilde{h}-\tilde{p}-\tilde{q})
    \ .
\end{equation}
Thus we have:
\begin{align}
    \delta\beta_B&=-6\beta\lambda^2 D\int_{\tilde{q}} (p_x+q_x)(h_x-p_x-q_x)G_0(\tilde{q})G_0(-\tilde{q})G_0(\tilde{p}+\tilde{q})G_0(\tilde{h}-\tilde{p}-\tilde{q}) \\
    &=6\beta\lambda^2 D\int_{\tilde{q}} q_x^2G_0^2(\tilde{q})G_0^2(-\tilde{q})
     \ ,
\end{align}
after going to the hydrodynamic limit $\tilde{p},\tilde{h}\rightarrow0$.

The C-graph yields a correction:
\begin{equation}
    \delta(\partial_t\phi)=12\beta D\lambda^2G_0(\tilde{k})\int_{\tilde{h},\tilde{p}}G_0(\tilde{k}-\tilde{h})G_0(\tilde{p})G_0(\tilde{h}-\tilde{p}) \int_{\tilde{q}} (p_x-q_x)(h_x-q_x)G_0(\tilde{q})G_0(-\tilde{q})G_0(\tilde{p}-\tilde{q})G_0(\tilde{h}-\tilde{q})
    \ .
\end{equation}
Thus:
\begin{align}
    \delta\beta_C&=-12\beta\lambda^2 D\int_{\tilde{q}} q_x^2G_0(\tilde{q})G_0^3(-\tilde{q})
    \ .
\end{align}
Using relations \ref{eq:res1} and \ref{eq:res2} we remind that:
\begin{align}
    &\int_{-\infty}^\infty \frac{dz}{2\pi} \frac{1}{(a-iz)^2(a+iz)^2}=\frac{1}{4a^3}
    \ ,
    \\
    &\int_{-\infty}^\infty \frac{dz}{2\pi} \frac{1}{(a-iz)^3(a+iz)}=\frac{1}{8a^3}
     \ .
\end{align}
Thus we indeed have after integration over $\Omega$:
\begin{equation}
    \delta\beta_B=-\delta\beta_C
    \ .
\end{equation}
As $\delta\beta=\delta\beta_A$ we finally have the flow equation for $\beta$:
\begin{equation}
    \frac{d\beta}{dl}=\beta\left(2\chi+z-\frac{27}{8\sqrt{2}}\frac{\beta D}{(\mu^7\nu)^{\frac{1}{4}}}\frac{S_{d-2}\Lambda^{d-4.5}}{(2\pi)^{d-1}} \right)
    \ .
\end{equation}

\subsubsection{Coupling constants and RG flow}
We now introduce two coupling constants $g_\lambda^x$ and $g_\beta^x$. $g_\lambda^x$ was already present at $d=5.5-\epsilon$, and will be diverging. $g_\beta^x$ is present because of the $\beta$ term, and will have a regular behavior. The definitions are:
\begin{align}
    g_\lambda^x&=\frac{D\lambda^2}{(\nu^3\mu^9)^\frac{1}{4}}\frac{S_{d-2}\Lambda^{d-5.5}}{(2\pi)^{d-1}}
    \ , \\
    g_\beta^x&=\frac{D\beta}{(\nu\mu^7)^\frac{1}{4}}\frac{S_{d-2}\Lambda^{d-4.5}}{(2\pi)^{d-1}}
    \ .
\end{align}
We can now rewrite the flow equations under the form:
\begin{align}
    \frac{d\mu}{dl}&=\mu(z-2)
     \ , \label{eq:flowLongiUncontrolledMu}\\
    \frac{d\nu}{dl}&=\nu\left(z-4\zeta -\frac{75}{1024\sqrt{2}}g_\lambda^x\right)  \ , \label{eq:flowLongiUncontrolledNu}\\
    \frac{d\lambda}{dl}&=\lambda\left(\chi+z-\zeta-\frac{63}{32\sqrt{2}}g_\beta^x \right) \ , \label{eq:flowLongiUncontrolledLambda}\\
    \frac{d\beta}{dl}&=\beta\left(2\chi+z-\frac{27}{8\sqrt{2}}g_\beta^x \right)  \ ,\label{eq:flowLongiUncontrolledBeta}\\
    \frac{dD}{dl}&=\frac{D}{2}(z-\zeta-2\alpha-d+1)
    \ . \label{eq:flowLongiUncontrolledD}
\end{align}
From the two linear flow equations we get the scaling relations:
\begin{align}
    z&=2 \ ,\\
    2\chi+\zeta&=3-d \ .
\end{align}
In order to obtain the RG flow in the coupling constants space $(g_\lambda,g_\beta)$, we once again log-differentiate the definitions of the coupling constants and inject the flow equations for the coefficients. We obtain the following dynamical system:
\begin{align}
    \frac{dg_\lambda^x}{dl}&= (5.5-d)g_\lambda^x +\frac{225}{4096\sqrt{2}}(g_\lambda^x)^2 - \frac{63}{16\sqrt{2}}g_\lambda^x g_\beta^x \ ,\\
    \frac{dg_\beta^x}{dl}&= (4.5-d)g_\beta^x -\frac{27}{8\sqrt{2}}(g_\beta^x)^2 + \frac{75}{4096\sqrt{2}}g_\lambda^x g_\beta^x  \ .
\end{align}
Although the RG flow seemingly depends on the spatial dimension, its general behavior doesn't change if $d<4.5$ (due to the presence of the $\beta$ term). Fig.2.(a) in the MT describes the flow: it diverges from any point where $g_\lambda$ is not zero. On the $g_\lambda=0$ axis however, there is a non-trivial fixed point. The set of all FPs is thus
\begin{align}
    (g_\lambda^{x,*},g_\beta^{x,*})_1&=(0,0)  \ ,\\
    (g_\lambda^{x,*},g_\beta^{x,*})_2&=\left(0,\frac{8\sqrt{2}}{27}(4.5-d)\right)
    \ .
\end{align}

\subsubsection{Anisotropic Ising Lifshitz universality class}
Focusing now on the only non-Gaussian FP, where $g_\lambda^{x,*}=0$ implying that $\lambda=0$, we see that the EOM reverts back to the equilibrium, albeit anisotropic, Ising Lifshitz UC. 
In order to obtain the universality class' exponents this FP, we need to inject the values of the coupling constants $(g_\lambda^{x,*},g_\beta^{x,*})_2$ into the flow equations \ref{eq:flowLongiUncontrolledMu}, \ref{eq:flowLongiUncontrolledNu}, \ref{eq:flowLongiUncontrolledBeta} and \ref{eq:flowLongiUncontrolledD}. Using three of the four remaining equations, we determine the scaling exponents $(\chi,\zeta,z)$, and check afterwards that the fourth equation is coherent. We get:
\begin{equation}
    \chi=\frac{5}{4}-\frac{d}{2} \sep \zeta=\frac{1}{2} \sep z=2 \ .
\end{equation}
These exponents being the same as those of the linear regime, we are led to think that the UC of this fixed point is the same as the linear regime's, although this equality is expected to change as one goes beyond 1-loop.

We can also note that the lower critical dimension is thus $2.5$ in the longitudinal Lifshitz point when fine-tuning $\lambda$ to zero. Evaluating them at $d=3$, we get:
\begin{equation}
    \chi_{d=3}=-\frac{1}{4} \sep \zeta=\frac{1}{2} \sep z=2 \ .
\end{equation}

\section{Transverse Lifshitz point}
We now focus on the Transverse Lifshitz point. To ease notation, we again
 use the notations $\mu$, $\nu$ instead of $\nu_\perp$, $\mu_x$.

\subsection{Linear theory}
\subsubsection{Linear exponents}
We study the transverse LP transition, governed by the equation of motion (11), and set the non-linear terms to zero for now. In Fourier space it can be re-written once again using equation \ref{eq:EOMlin}, this time with the linear propagator:
\begin{equation}
\label{eq:propTransverseLin}
    G_0(\tilde{k})=\frac{1}{\mu k_x^2+\nu k_\perp^4-i\omega}
    \ .
\end{equation}
This time the correlation function takes the form:
\begin{align}
    C_\phi(\mathbf{r},t)&=\int_{\tilde{k}} \frac{2De^{i(\mathbf{k\cdot r}-\omega t)}}{(\mu k_x^2 +\nu k_\perp^4-i\omega)(\mu k_x^2 +\nu k_\perp^4+i\omega)} \\
    &=|\mathbf{r_\perp}|^{3-d}\int_{\tilde{K}} \frac{2De^{i(\mathbf{K_\perp\cdot u} + K_x\frac{x}{|\mathbf{r_\perp}|^2}-\frac{\Omega t}{|\mathbf{r_\perp}|^4})}}{(\mu K_x^2 +\nu K_\perp^4-i\Omega)(\mu K_x^2 +\nu K_\perp^4+i\Omega)} \\
    &=|\mathbf{r_\perp}|^{3-d}S\left(\frac{x}{|\mathbf{r_\perp}|^2},\frac{t}{|\mathbf{r_\perp}|^4} \right)
    \ .
\end{align}
Which gives us the linear exponents (12):
\begin{equation}
\label{eq:CriticalLinExpTrans}
    \chi_T^{\rm lin}=\frac{3}{2}-\frac{d}{2} \sep \zeta_T^{\rm lin}=2 \sep z_T^{\rm lin}=4
    \ .
\end{equation}

\subsubsection{Upper critical dimension}
From these we can extract the upper critical dimension of the transverse LP, using power counting. We get after rescaling by a factor $a$, $\mathbf{r_\perp} \longrightarrow a\mathbf{r_\perp}$ :
\begin{equation}
    \partial_t\phi= \mu\partial_x^2\phi - \nu\nabla_\perp^4\phi + \frac{\lambda}{2}a^{\chi+z-\zeta}\partial_x\phi^2 -\beta a^{2\chi+z}\phi^3+f
    \ ,
\end{equation}
yielding the followings:
\begin{itemize}
\item The $\lambda$ term is rescaled by a factor $a^{\frac{7-d}{2}}$, meaning that its upper critical dimension is $d_c=7$. 
\item The $\beta$ term is rescaled by a factor $a^{7-d}$, meaning that its upper critical dimension is also $d_c=7$. 
\end{itemize}
Contrary to the longitudinal LP, both non-linear terms have to be simultaneously renormalized in the DRG analysis, which is applied with $\epsilon=7-d$.

\subsection{Dynamical Renormalization Group analysis in dimension $7-\epsilon$}
\subsubsection{EOM in Fourier space and diagrammatic expansion}
In Fourier space the EOM writes:
\begin{equation}
    \phi(\tilde{k}) = G_0(\tilde{k})\eta(\tilde{k}) + i\frac{\lambda}{2}k_xG_0(\tilde{k})\int_{\tilde{q}}\phi(\tilde{q})\phi(\tilde{k}-\tilde{q}) -\beta G_0(\tilde{k})\int_{\tilde{p}_1,\tilde{p}_2}\phi(\tilde{p}_2)\phi(\tilde{p}_1-\tilde{p}_2)\phi(\tilde{k}-\tilde{p}_1)
    \ .
\end{equation}
with the new definition \ref{eq:propTransverseLin} of $G_0$.

Because the EOM in the transverse LP has the same structure as the one in fixed dimension $d<4.5$ of the longitudinal LP (only $G_0$ is different), the diagrammatic expansion is the same. We thus refer to expansions \ref{eq:prop1}, \ref{eq:noisediagram}, \ref{eq:vertex3diagramUncontrolled} and \ref{eq:vertex4diagramUncontrolled} respectively for the propagator, noise, $\lambda$-vertex and $\beta$-vertex. We now re-derive the flow equations.

\subsubsection{Propagator flow equations}
Referring to equation \ref{eq:prop1}, one yields:
\begin{equation}
\label{eq:propExpansionTransverse}
    \delta(\partial_t\phi)=-\frac{1}{2}D\lambda^2k_xG_0^2(\tilde{k})\int_{\tilde{q}} (\frac{k_x}{2}-q_x)G_0(\frac{\tilde{k}}{2}-\tilde{q})G_0(\tilde{q}-\frac{\tilde{k}}{2})G_0(\frac{\tilde{k}}{2}+\tilde{q})
     \ .
\end{equation}

Evaluating the propagator expansion in $k_x=0,\omega=0$ we yield:
\begin{equation}
    \frac{1}{\nu^<k_\perp^4}=\frac{1}{\nu k_\perp^4}
    \ .
\end{equation}
Thus $\nu^<=\nu$, meaning that there is no non-linear correction and we thus have the flow equation for $\mu$:
\begin{equation}
    \frac{d\nu}{dl}=\nu \left( z-4 \right)
    \ .
\end{equation}

In order to get the correction for $\mu$ we evaluate \ref{eq:propExpansionTransverse} at $\mathbf{k_\perp}=\mathbf{0},\omega=0$ and need to expand the integral to first order in $k_x$ (as $G_0^2$ corresponds to $k_x^{-2}$). Expanding the three $G_0$ terms inside the integral, then using again the residue theorem relations \ref{eq:res1}-\ref{eq:res4}, we get:
\begin{align}
    \delta(\partial_t\phi)&=-2D\lambda^2k_x^2G_0^2(\tilde{k})\int_{\tilde{q}} \left( \frac{1}{2}G_0(\tilde{q})G_0^2(-\tilde{q}) + 2\mu q_x^2G_0(\tilde{q})G_0^3(-\tilde{q}) -\mu q_x^2G_0^2(\tilde{q})G_0^2(-\tilde{q}) \right) \\
    &=-2D\lambda^2k_x^2G_0^2(\tilde{k})\times\frac{1}{32}\frac{1}{(\nu^3\mu)^{\frac{1}{2}}} \int_{\mathbf{q_\perp}}^\Lambda \frac{1}{q_\perp^6}
    \ .
\end{align}
Finally we get the flow equation:
\begin{equation}
    \frac{d\mu}{dl}=\mu \left( z-2\zeta +\frac{1}{16}\frac{D\lambda^2}{(\nu\mu)^{\frac{3}{2}}}\frac{S_{d-2}\Lambda^{d-7}}{(2\pi)^{d-1}} \right)
    \ .
\end{equation}

\subsubsection{Noise flow equation}
Just like in the previous cases, the noise cannot be renormalized, and the flow equation is simply the linear case.

\subsubsection{$\lambda$-vertex flow equation}
Just like in the longitudinal LP, the three $\Gamma_a$,$\Gamma_b$ and $\Gamma_c$ terms cancel each other. This comes essentially from the fact that the cancellation occurs at the $\Omega$ integration, which doesn't use the full definition of $G_0$. The part which we call $a$ is considered as a constant, so the integrals cancel no matter the value of $a$, and thus the cancellation of the graphs in this case too.

The A-graph term from \ref{eq:vertex3diagramUncontrolled} leads to a correction:
\begin{align}
    \delta(\partial_t\phi)&=-6iD\lambda\beta G_0(\tilde{k})k_x\int_{\tilde{h}}G_0(\tilde{h})G_0(\tilde{k}-\tilde{h}) \int_{\tilde{q}} G_0(\tilde{q})G_0(-\tilde{q})G_0(\tilde{k}-\tilde{q}) 
    \ ,\\
    \delta\lambda_A&=-12\beta D\lambda \int_{\tilde{q}} G_0(\tilde{q})G_0^2(-\tilde{q}) \\
    &=-\frac{3}{4}\frac{\beta D\lambda}{(\nu^3\mu)^{\frac{1}{2}}}\int_{\mathbf{q_\perp}}^{\Lambda} \frac{1}{q_\perp^6} \\
    &=-\frac{3}{4}\frac{\beta D\lambda}{(\nu^3\mu)^{\frac{1}{2}}}\frac{S_{d-2}\Lambda^{d-7}}{(2\pi)^{d-1}}dl
    \ .
\end{align}

The B-graph corrections write, using the same techniques ($G_0$ expansions in the internal integral, residue calculations, integrations):
\begin{align}
    \delta(\partial_t\phi)&=-12iD\lambda\beta G_0(\tilde{k})\int_{\tilde{h}}G_0(\tilde{h})G_0(\tilde{k}-\tilde{h}) \int_{\tilde{q}} (\frac{h_x}{2}+q_x)G_0(\frac{\tilde{h}}{2}+\tilde{q})G_0(\frac{\tilde{h}}{2}-\tilde{q})G_0(\tilde{q}-\frac{\tilde{h}}{2})\ , \\
    \delta\lambda_B&=-\frac{3}{8}\frac{\beta D\lambda}{(\nu^3\mu)^{\frac{1}{2}}}\frac{S_{d-2}\Lambda^{d-7}}{(2\pi)^{d-1}}dl\ .
\end{align}

We thus finally get the $\lambda$ flow equation:
\begin{equation}
    \frac{d\lambda}{dl}=\lambda\left(\chi+z-\zeta -\frac{9}{8}\frac{\beta D}{(\nu^3\mu)^{\frac{1}{2}}}\frac{S_{d-2}\Lambda^{d-7}}{(2\pi)^{d-1}} \right)
    \ .
\end{equation}

\subsubsection{$\beta$-vertex flow equation}
Here we are once more using equation \ref{eq:vertex4diagramUncontrolled}, and recall that the B- and C-graphs cancel each other. Thus the only correction comes from the A-graph, which yields:
\begin{align}
    \delta(\partial_t\phi)&=36\beta^2DG_0(\tilde{k})\int_{\tilde{h},\tilde{p}}G_0(\tilde{k}-\tilde{h})G_0(\tilde{p})G_0(\tilde{h}-\tilde{p}) \int_{\tilde{q}} G_0(\tilde{q})G_0(-\tilde{q})G_0(\tilde{h}-\tilde{q}) 
    \ , \\
    \delta\beta&=-36\beta^2D\int_{\tilde{q}}G_0(\tilde{q})G_0^2(-\tilde{q}) \\
  &=-\frac{9}{4}\frac{\beta^2D}{(\mu\nu^3)^{\frac{1}{2}}}\frac{S_{d-2}\Lambda^{d-7}}{(2\pi)^{d-1}}dl
  \ .
\end{align}
And we yield the $\beta$ flow equation:
\begin{equation}
    \frac{d\beta}{dl}=\beta\left(2\chi+z -\frac{9}{4}\frac{\beta^2D}{(\mu\nu^3)^{\frac{1}{2}}}\frac{S_{d-2}\Lambda^{d-7}}{(2\pi)^{d-1}} \right)\ .
\end{equation}

\subsubsection{RG flow in the transverse Lifshitz point}
We now introduce the two natural coupling constants $g_\lambda^\perp$ and $g_\beta^\perp$ in the transverse Lifshitz point
\beq
    g_\lambda^\perp=\frac{D\lambda^2}{(\mu\nu)^\frac{3}{2}}\frac{S_{d-2}\Lambda^{d-7}}{(2\pi)^{d-1}} \sep
    g_\beta^\perp=\frac{D\beta}{(\nu^3\mu)^\frac{1}{2}}\frac{S_{d-2}\Lambda^{d-7}}{(2\pi)^{d-1}}
    \ ,
\eeq
 and re-write the flow equations :
\begin{align}
    \frac{d\nu}{dl}&=\nu \left( z-2\zeta +\frac{1}{16}g_\lambda^\perp \right)\ , \label{eq:flowTransNu}\\
    \frac{d\mu}{dl}&=\mu \left( z-4 \right)\ ,\label{eq:flowTransMu}\\
    \frac{d\lambda}{dl}&=\lambda\left(\chi+z-\zeta -\frac{9}{8}g_\beta^\perp \right)\ , \label{eq:flowTransLambda}\\
    \frac{d\beta}{dl}&=\beta\left(2\chi+z -\frac{9}{4}g_\beta^\perp \right)\ , \label{eq:flowTransBeta}\\
    \frac{dD}{dl}&=\frac{D}{2}(z-\zeta-2\alpha-d+1)
    \ . \label{eq:flowTransD}
\end{align}
Once again by log-differentiating the definition of coupling constants and injecting the flow equations, we get the dynamical system in the coupling constants space:
\begin{align}
    \frac{dg_\lambda^\perp}{dl}&= \epsilon g_\lambda^\perp -\frac{3}{32}(g_\lambda^\perp)^2 - \frac{9}{4}g_\lambda^\perp g_\beta^\perp \ , \\
    \frac{dg_\beta^\perp}{dl}&= \epsilon g_\beta^\perp -\frac{9}{4}(g_\beta^\perp)^2 - \frac{1}{32}g_\lambda^\perp g_\beta^\perp \ ,
\end{align}
where we use $\epsilon=7-d$. The flow is represented in Fig.2.(b), and contrary to the longitudinal case, has a converging behavior. There are three fixed points worth noticing, including the linear regime FP, a generic linear-like FP, and a non-generic FP which behaves according to a new universality class:
\begin{align}
    (g_\lambda^{\perp,*},g_\beta^{\perp,*})_{linear}&=(0,0)  \ ,\\
    (g_\lambda^{\perp,*},g_\beta{\perp,^*})_{generic}&=(0,\frac{4}{9}\epsilon) \ ,\\
    (g_\lambda^{\perp,*},g_\beta^{\perp,*})_{new\;UC}&=(\frac{32}{2}\epsilon,0)\ .
\end{align}
Here, the ``generic" fixed point corresponds again to the equilibrium, anisotropic Ising Lifshitz point.

\subsubsection{Universality classes}
The critical exponents of the linear regime of the transverse Lifshitz point are given by \ref{eq:CriticalLinExpTrans} or (12). The critical exponents of the generic FP are given by (after injecting the coupling constants values in the flow equations):
\begin{equation}
    \chi_{generic}=\frac{\epsilon}{2}-2 \sep \zeta_{generic}=2 \sep z_{generic}=4
    \ .
\end{equation}
These are similar to the linear regime's, which can be quickly seen, as when $g_\lambda=0$, equation \ref{eq:flowTransNu} turns back to the linear regime, which along with \ref{eq:flowTransMu} and \ref{eq:flowTransD} gives three equations similar to the linear regime. The interesting fact, however, is that equation \ref{eq:flowTransBeta} agrees with these values. The lower critical dimension of this set of exponents is $d_{lower}=3$, meaning that the behavior in dimension 3 is not exactly known. We are however not sure, once more, if the UC is actually the same as the linear regime, or if this is due to the 1-loop approximation.

However, when fine-tuning $\beta$ to zero, we get the fixed point $(g_\lambda^*,g_\beta^*)_{new\;UC}$, which has a different UC:
\begin{equation}
    \chi_{new}=\frac{\epsilon}{3}-2 \sep \zeta_{new}=\frac{\epsilon}{3}+2 \sep z_{new}=4
    \ .
\end{equation}
which is so far unseen to us. Furthermore, this set of exponents has a lower critical dimension $d_{lower}=1$, meaning that this allows us to obtain an approximation in dimension 3:
\begin{equation}
    \chi_{new,d=3}=-\frac{2}{3} \sep \zeta_{new,d=3}=\frac{10}{3} \sep z_{new,d=3}=4
    \ .
\end{equation}

\section{Simultaneous Longitudinal and Transverse Lifshitz Point}
For completeness, we now consider the case of simultaneously fine-tuning $\mu_x$ and $\mu_\perp$ to zero in our AMIM.
\subsection{Linear theory}
\subsubsection{Linear correlation function and linear exponents}
This case corresponds to the linear propagator:
\begin{equation}
\label{eq:proplinsimu}
    G_0(\tilde{k})=\frac{1}{\nu_xk_x^4 +\nu_\perp^4k_\perp^4-i\omega}
    \ ,
\end{equation}
leading to the correlation function:
\begin{align}
    C_\phi(\mathbf{r},t)&=\int_{\tilde{k}} \frac{2De^{i(\mathbf{k\cdot r}-\omega t)}}{(\nu_x k_x^4 +\nu k_\perp^4-i\omega)(\nu_x k_x^4 +\nu k_\perp^4+i\omega)}
    \nonumber\\
    &=|\mathbf{r_\perp}|^{4-d}\int_{\tilde{K}} \frac{2De^{i(\mathbf{K_\perp\cdot u} + K_x\frac{x}{|\mathbf{r_\perp}|}-\frac{\Omega t}{|\mathbf{r_\perp}|^4})}}{(\nu_x K_x^4 +\nu K_\perp^4-i\Omega)(\nu_x K_x^4 +\nu K_\perp^4+i\Omega)} \nonumber\\
    &=|\mathbf{r_\perp}|^{4-d}f\left(\frac{x}{|\mathbf{r_\perp}|},\frac{t}{|\mathbf{r_\perp}|^4} \right)
    \ ,
\end{align}
which gives us the linear exponents:
\begin{equation}
\label{eq:criticalLinExponentsSimu}
    \chi^{\rm lin}=2-\frac{d}{2} \sep \zeta^{\rm lin}=1 \sep z^{\rm lin}=4
    \ .
\end{equation}

\subsubsection{Upper critical dimension}
After rescaling by a factor $a$, $\mathbf{r_\perp} \longrightarrow a\mathbf{r_\perp}$ we have at the linear level:
\begin{equation}
    \partial_t\phi= \nu_x\partial_x^2\phi - \nu_\perp\nabla_\perp^4\phi + \frac{\lambda}{2}a^{\chi+z-\zeta}\partial_x\phi^2 -\beta a^{2\chi+z}\phi^3+f
    \ .
\end{equation}
We can thus conclude the followings:
\begin{itemize}
    \item The $\lambda$ term is rescaled by a factor $a^{(\chi-\zeta+z)_{lin}}=a^{5-\frac{d}{2}}$, thus its upper critical dimension is $d_c=10$.
    \item The $\beta$ term is rescaled by a factor $a^{(2\chi+z)_{lin}}=a^{8-d}$, thus its upper critical dimension is $d_c=8$.
\end{itemize}
In the $\epsilon$-expansion we perform next, the $\beta$ term is thus irrelevant.

\subsection{DRG analysis in dimension $10-\epsilon$}
\subsubsection{EOM in Fourier space and diagrammatic expansions}
In this approximation we remark that the EOM in Fourier space writes exactly in the same way as in the longitudinal LP case \ref{eq:EOMLongi}:
\begin{equation}
    \phi(\tilde{k}) = G_0(\tilde{k})\eta(\tilde{k}) + i\frac{\lambda}{2}k_xG_0(\tilde{k})\int_{\tilde{q}}\phi(\tilde{q})\phi(\tilde{k}-\tilde{q})
    \ ,
\end{equation}
with the only difference that the linear propagator $G_0$ takes the form \ref{eq:proplinsimu}. This similarity will allow us to skip many steps in the calculation and we will eventually yield a similar result, which is that the first order phase transition extends to this ``all directions" LP.

In particular the diagrammatic expansions up to one loop are exactly \ref{eq:prop1}-\ref{eq:noisediagram}.

\subsubsection{Propagator expansion}
The diagrammatic expansion is given by \ref{eq:prop1}. The associated equation is: 
\begin{equation}
    G(\tilde{k}) = G_0(\tilde{k}) - 2D\lambda^2G_0^2(\tilde{k})k_x\int_{\tilde{q}}(\frac{k_x}{2}+q_x)G_0(\frac{\tilde{k}}{2}+\tilde{q})G_0(\frac{\tilde{k}}{2}-\tilde{q})G_0(\tilde{q}-\frac{\tilde{k}}{2}) \; + O(\lambda^4)
    \ .
\end{equation}

Evaluating this equation in $\tilde{k}=(k_x=0,\mathbf{k_\perp},\omega=0)$, we yield no correction to the linear regime and yield the first flow equation:
\begin{equation}
\label{eq:flowSimuNuPerp}
    \frac{d\nu_\perp}{dl}=\nu_\perp\left(z-4\right)
    \ .
\end{equation}

Evaluation the propagator expansion in $\tilde{k}=(k_x,\mathbf{k_\perp}=\mathbf{0},\omega=0)$, we will get a correction for $\nu_x$. For this we first expand the integral to third order in $k_x$ using series expansions of $G_0(\frac{\tilde{k}}{2}+\tilde{q})$. Note that it will in fact be the same as in the longitudinal LP case because the dependence of $G_0$ in $k_x$ is the same:
\begin{align}
    G_0\left(\frac{\tilde{k}}{2}+\tilde{q}\right)&=\frac{1}{\nu_\perp q_\perp^4+\nu_x(\frac{k_x}{2}+q_x)^4-i\Omega}+O(k_x^4)\nonumber\\
    &=\frac{1}{G_0(\tilde{q})\left(1+\frac{\nu_x}{G_0(\tilde{q})}(2q_x^3k_x+\frac{3}{2}q_x^2k_x^2+\frac{1}{2}q_xk_x^3)\right)}+O(k_x^4)
    \ ,
\end{align}
which exactly leads to equation \ref{eq:linpropexp} for which $G_0(\tilde{q})$ has the new definition. The rest of the calculation is thus formally equivalent as we integrate over $\Omega$ by introducing the constant $a$ which takes the new definition $a=\nu_xq_x^4+\nu_\perp q_\perp^4$, yielding
\begin{align}
    \int_{\tilde{q}}(\frac{k_x}{2}+q_x)G_0(\frac{\tilde{k}}{2}+\tilde{q})G_0(\frac{\tilde{k}}{2}-\tilde{q})G_0(\tilde{q}-\frac{\tilde{k}}{2}) &= -\frac{15}{32}\nu_xk_x^3\int_q\frac{q_x^2}{(\nu_\perp q_\perp^4+\nu_xq_x^4)^3}\; +O(k_x^4) \\
    &=-\frac{15}{32}\nu_xk_x^3\int_{q_\perp}^\Lambda \frac{1}{(\nu_\perp^3\nu_x)^\frac{3}{4}q_\perp^9}\frac{1}{2\pi}\int_{-\infty}^{\infty}\frac{u^2}{(1+u^4)^3}du \; +O(k_x^4) \\
    &=-\frac{75}{2048\sqrt{2}}\frac{\nu_xk_x^3}{(\nu_\perp^3\nu_x)^\frac{3}{4}}\int_{q_\perp}^\Lambda \frac{1}{q_\perp^9} \; +O(k_x^4)
    \ .
\end{align}
This leads to a similar expression to \ref{eq:NuSupexp} but with a new definition of the coupling constant:
\begin{equation}
    \nu_x^<=\nu_x\left[1 -\frac{75}{1024\sqrt{2}}\frac{D\lambda^2}{(\nu_\perp^3\nu_x)^\frac{3}{4}}\frac{S_{d-2}}{(2\pi)^{d-1}}\Lambda^{d-10}dl\right]
    \ .
\end{equation}
And we get the second flow equation:
\begin{equation}
\label{eq:flowSimuNuX}
    \frac{d\nu_x}{dl}=\nu_x\left(z-4\zeta -\frac{75}{1024\sqrt{2}}\frac{D\lambda^2}{(\nu_\perp^3\nu_x)^\frac{3}{4}}\frac{S_{d-2}}{(2\pi)^{d-1}}\Lambda^{d-10} \right)
    \ .
\end{equation}

\subsubsection{$\lambda$ vertex expansion}
The diagrammatic expansion is \ref{eq:vertexdiagram1}, for which the corrections are once again equal to zero. Indeed we write it in the same form as \ref{eq:gammaA}-\ref{eq:gammaC}, simply using the new definition of $G_0$. Because we only need the zeroth order in $k_x$, this doesn't create any different term that could appear in an series expansion. Then the cancellation of the three terms comes from the integration over $\Omega$, and the modifications on $G_0$ don't change anything.

Thus $\lambda$ doesn't get any corrections, yielding the previous flow equation:
\begin{equation}
\label{eq:flowSimuLambda}
    \frac{d\lambda}{dl}=\lambda\left(\chi-\zeta+z \right)\ .
\end{equation}

\subsubsection{Noise expansion}
The diagrammatic expansion is still \ref{eq:noisediagram} and, as in all the previous cases, the analytic expression \ref{eq:noisesupexp} doesn't yield any correction as it goes to zero in the hydrodynamic limit $k_x \longrightarrow 0$. 

Thus:
\begin{equation}
\label{eq:flowSimuD}
    \frac{dD}{dl}=\frac{D}{2}\left(-2\chi-\zeta+z-d+1 \right)\ .
\end{equation}

\subsection{RG flow}
The system of flow equations \ref{eq:flowSimuNuPerp}, \ref{eq:flowSimuNuX}-\ref{eq:flowSimuD} has the exact same structure as that of the longitudinal LP (up to numerical coefficients), with a different definition of the coupling constant:
\begin{equation}
    g_\lambda^{x+\perp}=\frac{D\lambda^2}{(\nu_\perp^3\nu_x)^{\frac{3}{4}}}\frac{S_{d-2}\Lambda^{d-10}}{(2\pi)^{d-1}}
    \ .
\end{equation}
Log-differentiating this definition like in previous cases, we yield the flow dynamics:
\begin{equation}
    \frac{dg_\lambda^{x+\perp}}{dl}=\epsilon g_\lambda^{x+\perp} + \frac{225}{4096\sqrt{2}}(g_\lambda^{x+\perp})^2
    \ .
\end{equation}

This is exactly the same equation as in the $5.5-\epsilon$ expansion of the longitudinal LP, meaning that the diverging behavior of the flow is the same. The only difference here is that the linear regime $\lambda=0$ is characterized by a different set of linear critical exponents \ref{eq:criticalLinExponentsSimu}.

We thus conclude that the first-type phase transition of the longitudinal LP extends to the case of the ``all directions" LP.


\begin{thebibliography}{22}%
\makeatletter
\providecommand \@ifxundefined [1]{%
 \@ifx{#1\undefined}
}%
\providecommand \@ifnum [1]{%
 \ifnum #1\expandafter \@firstoftwo
 \else \expandafter \@secondoftwo
 \fi
}%
\providecommand \@ifx [1]{%
 \ifx #1\expandafter \@firstoftwo
 \else \expandafter \@secondoftwo
 \fi
}%
\providecommand \natexlab [1]{#1}%
\providecommand \enquote  [1]{``#1''}%
\providecommand \bibnamefont  [1]{#1}%
\providecommand \bibfnamefont [1]{#1}%
\providecommand \citenamefont [1]{#1}%
\providecommand \href@noop [0]{\@secondoftwo}%
\providecommand \href [0]{\begingroup \@sanitize@url \@href}%
\providecommand \@href[1]{\@@startlink{#1}\@@href}%
\providecommand \@@href[1]{\endgroup#1\@@endlink}%
\providecommand \@sanitize@url [0]{\catcode `\\12\catcode `\$12\catcode
  `\&12\catcode `\#12\catcode `\^12\catcode `\_12\catcode `\%12\relax}%
\providecommand \@@startlink[1]{}%
\providecommand \@@endlink[0]{}%
\providecommand \url  [0]{\begingroup\@sanitize@url \@url }%
\providecommand \@url [1]{\endgroup\@href {#1}{\urlprefix }}%
\providecommand \urlprefix  [0]{URL }%
\providecommand \Eprint [0]{\href }%
\providecommand \doibase [0]{https://doi.org/}%
\providecommand \selectlanguage [0]{\@gobble}%
\providecommand \bibinfo  [0]{\@secondoftwo}%
\providecommand \bibfield  [0]{\@secondoftwo}%
\providecommand \translation [1]{[#1]}%
\providecommand \BibitemOpen [0]{}%
\providecommand \bibitemStop [0]{}%
\providecommand \bibitemNoStop [0]{.\EOS\space}%
\providecommand \EOS [0]{\spacefactor3000\relax}%
\providecommand \BibitemShut  [1]{\csname bibitem#1\endcsname}%
\let\auto@bib@innerbib\@empty
\bibitem [{\citenamefont {Hornreich}\ \emph {et~al.}(1975)\citenamefont
  {Hornreich}, \citenamefont {Luban},\ and\ \citenamefont
  {Shtrikman}}]{hornreich_prl75}%
  \BibitemOpen
  \bibfield  {author} {\bibinfo {author} {\bibfnamefont {R.~M.}\ \bibnamefont
  {Hornreich}}, \bibinfo {author} {\bibfnamefont {M.}~\bibnamefont {Luban}},\
  and\ \bibinfo {author} {\bibfnamefont {S.}~\bibnamefont {Shtrikman}},\
  }\bibfield  {title} {\bibinfo {title} {Critical {Behavior} at the {Onset} of
  k-{Space} {Instability} on the $\lambda$ {Line}},\ }\href
  {https://doi.org/10.1103/PhysRevLett.35.1678} {\bibfield  {journal} {\bibinfo
   {journal} {Physical Review Letters}\ }\textbf {\bibinfo {volume} {35}},\
  \bibinfo {pages} {1678} (\bibinfo {year} {1975})}\BibitemShut {NoStop}%
\bibitem [{\citenamefont {Hornreich}\ \emph {et~al.}(1979)\citenamefont
  {Hornreich}, \citenamefont {Liebmann}, \citenamefont {Schuster},\ and\
  \citenamefont {Selke}}]{hornreich_zpb79}%
  \BibitemOpen
  \bibfield  {author} {\bibinfo {author} {\bibfnamefont {R.~M.}\ \bibnamefont
  {Hornreich}}, \bibinfo {author} {\bibfnamefont {R.}~\bibnamefont {Liebmann}},
  \bibinfo {author} {\bibfnamefont {H.~G.}\ \bibnamefont {Schuster}},\ and\
  \bibinfo {author} {\bibfnamefont {W.}~\bibnamefont {Selke}},\ }\bibfield
  {title} {\bibinfo {title} {Lifshitz points in ising systems},\ }\href
  {https://doi.org/10.1007/BF01322086} {\bibfield  {journal} {\bibinfo
  {journal} {Zeitschrift für Physik B Condensed Matter}\ }\textbf {\bibinfo
  {volume} {35}},\ \bibinfo {pages} {91} (\bibinfo {year} {1979})}\BibitemShut {NoStop}%
\bibitem [{\citenamefont {Michelson}(1977)}]{michelson_prl77}%
  \BibitemOpen
  \bibfield  {author} {\bibinfo {author} {\bibfnamefont {A.}~\bibnamefont
  {Michelson}},\ }\bibfield  {title} {\bibinfo {title} {Physical {Realization}
  of a {Lifshitz} {Point} in {Liquid} {Crystals}},\ }\href
  {https://doi.org/10.1103/PhysRevLett.39.464} {\bibfield  {journal} {\bibinfo
  {journal} {Physical Review Letters}\ }\textbf {\bibinfo {volume} {39}},\
  \bibinfo {pages} {464} (\bibinfo {year} {1977})}\BibitemShut {NoStop}%
\bibitem [{\citenamefont {Hořava}(2009)}]{horava_prd09}%
  \BibitemOpen
  \bibfield  {author} {\bibinfo {author} {\bibfnamefont {P.}~\bibnamefont
  {Hořava}},\ }\bibfield  {title} {\bibinfo {title} {Quantum gravity at a
  {Lifshitz} point},\ }\href {https://doi.org/10.1103/PhysRevD.79.084008}
  {\bibfield  {journal} {\bibinfo  {journal} {Physical Review D}\ }\textbf
  {\bibinfo {volume} {79}},\ \bibinfo {pages} {084008} (\bibinfo {year}
  {2009})}\BibitemShut {NoStop}%
\bibitem [{\citenamefont {Ramaswamy}(2010)}]{ramaswamy_annrev10}%
  \BibitemOpen
  \bibfield  {author} {\bibinfo {author} {\bibfnamefont {S.}~\bibnamefont
  {Ramaswamy}},\ }\bibfield  {title} {\bibinfo {title} {The {Mechanics} and
  {Statistics} of {Active} {Matter}},\ }\href
  {https://doi.org/10.1146/annurev-conmatphys-070909-104101} {\bibfield
  {journal} {\bibinfo  {journal} {Annual Review of Condensed Matter Physics}\
  }\textbf {\bibinfo {volume} {1}},\ \bibinfo {pages} {323} (\bibinfo {year}
  {2010})}\BibitemShut {NoStop}%
\bibitem [{\citenamefont {Marchetti}\ \emph {et~al.}(2013)\citenamefont
  {Marchetti}, \citenamefont {Joanny}, \citenamefont {Ramaswamy}, \citenamefont
  {Liverpool}, \citenamefont {Prost}, \citenamefont {Rao},\ and\ \citenamefont
  {Simha}}]{marchetti_rmp13}%
  \BibitemOpen
  \bibfield  {author} {\bibinfo {author} {\bibfnamefont {M.~C.}\ \bibnamefont
  {Marchetti}}, \bibinfo {author} {\bibfnamefont {J.~F.}\ \bibnamefont
  {Joanny}}, \bibinfo {author} {\bibfnamefont {S.}~\bibnamefont {Ramaswamy}},
  \bibinfo {author} {\bibfnamefont {T.~B.}\ \bibnamefont {Liverpool}}, \bibinfo
  {author} {\bibfnamefont {J.}~\bibnamefont {Prost}}, \bibinfo {author}
  {\bibfnamefont {M.}~\bibnamefont {Rao}},\ and\ \bibinfo {author}
  {\bibfnamefont {R.~A.}\ \bibnamefont {Simha}},\ }\bibfield  {title} {\bibinfo
  {title} {Hydrodynamics of soft active matter},\ }\href
  {https://doi.org/10.1103/RevModPhys.85.1143} {\bibfield  {journal} {\bibinfo
  {journal} {Reviews of Modern Physics}\ }\textbf {\bibinfo {volume} {85}},\
  \bibinfo {pages} {1143} (\bibinfo {year} {2013})}\BibitemShut {NoStop}%
\bibitem [{\citenamefont {Hohenberg}\ and\ \citenamefont
  {Halperin}(1977)}]{hohenberg_rmp77}%
  \BibitemOpen
  \bibfield  {author} {\bibinfo {author} {\bibfnamefont {P.~C.}\ \bibnamefont
  {Hohenberg}}\ and\ \bibinfo {author} {\bibfnamefont {B.~I.}\ \bibnamefont
  {Halperin}},\ }\bibfield  {title} {\bibinfo {title} {Theory of dynamic
  critical phenomena},\ }\href {https://doi.org/10.1103/RevModPhys.49.435}
  {\bibfield  {journal} {\bibinfo  {journal} {Reviews of Modern Physics}\
  }\textbf {\bibinfo {volume} {49}},\ \bibinfo {pages} {435} (\bibinfo {year}
  {1977})}\BibitemShut
  {NoStop}%
\bibitem [{\citenamefont {Solon}\ and\ \citenamefont
  {Tailleur}(2013)}]{solon_prl13}%
  \BibitemOpen
  \bibfield  {author} {\bibinfo {author} {\bibfnamefont {A.~P.}\ \bibnamefont
  {Solon}}\ and\ \bibinfo {author} {\bibfnamefont {J.}~\bibnamefont
  {Tailleur}},\ }\bibfield  {title} {\bibinfo {title} {Revisiting the flocking
  transition using active spins.},\ }\href
  {https://doi.org/10.1103/PhysRevLett.111.078101} {\bibfield  {journal}
  {\bibinfo  {journal} {Physical review letters}\ }\textbf {\bibinfo {volume}
  {111}},\ \bibinfo {pages} {078101} (\bibinfo {year} {2013})}\BibitemShut {NoStop}%
\bibitem [{\citenamefont {Solon}\ and\ \citenamefont
  {Tailleur}(2015)}]{solon_pre15}%
  \BibitemOpen
  \bibfield  {author} {\bibinfo {author} {\bibfnamefont {A.~P.}\ \bibnamefont
  {Solon}}\ and\ \bibinfo {author} {\bibfnamefont {J.}~\bibnamefont
  {Tailleur}},\ }\bibfield  {title} {\bibinfo {title} {Flocking with discrete
  symmetry: {The} two-dimensional active {Ising} model.},\ }\href
  {https://doi.org/10.1103/PhysRevE.92.042119} {\bibfield  {journal} {\bibinfo
  {journal} {Physical review. E, Statistical, nonlinear, and soft matter
  physics}\ }\textbf {\bibinfo {volume} {92}},\ \bibinfo {pages} {042119}
  (\bibinfo {year} {2015})}\BibitemShut {NoStop}%
\bibitem [{\citenamefont {Toner}(2012)}]{toner_prl12}%
  \BibitemOpen
  \bibfield  {author} {\bibinfo {author} {\bibfnamefont {J.}~\bibnamefont
  {Toner}},\ }\bibfield  {title} {\bibinfo {title} {Birth, {Death}, and
  {Flight}: {A} {Theory} of {Malthusian} {Flocks}},\ }\href
  {https://doi.org/10.1103/PhysRevLett.108.088102} {\bibfield  {journal}
  {\bibinfo  {journal} {Physical Review Letters}\ }\textbf {\bibinfo {volume}
  {108}},\ \bibinfo {pages} {088102} (\bibinfo {year} {2012})}\BibitemShut {NoStop}%
\bibitem [{\citenamefont {Chen}\ \emph
  {et~al.}(2020{\natexlab{a}})\citenamefont {Chen}, \citenamefont {Lee},\ and\
  \citenamefont {Toner}}]{chen_prl20}%
  \BibitemOpen
  \bibfield  {author} {\bibinfo {author} {\bibfnamefont {L.}~\bibnamefont
  {Chen}}, \bibinfo {author} {\bibfnamefont {C.~F.}\ \bibnamefont {Lee}},\ and\
  \bibinfo {author} {\bibfnamefont {J.}~\bibnamefont {Toner}},\ }\bibfield
  {title} {\bibinfo {title} {Moving, {Reproducing}, and {Dying} {Beyond}
  {Flatland}: {Malthusian} {Flocks} in {Dimensions} d {\textgreater} 2},\
  }\href {https://doi.org/10.1103/PhysRevLett.125.098003} {\bibfield  {journal}
  {\bibinfo  {journal} {Physical Review Letters}\ }\textbf {\bibinfo {volume}
  {125}},\ \bibinfo {pages} {098003} (\bibinfo {year} {2020}{\natexlab{a}})}\BibitemShut {NoStop}%
\bibitem [{\citenamefont {Chen}\ \emph
  {et~al.}(2020{\natexlab{b}})\citenamefont {Chen}, \citenamefont {Lee},\ and\
  \citenamefont {Toner}}]{chen_pre20}%
  \BibitemOpen
  \bibfield  {author} {\bibinfo {author} {\bibfnamefont {L.}~\bibnamefont
  {Chen}}, \bibinfo {author} {\bibfnamefont {C.~F.}\ \bibnamefont {Lee}},\ and\
  \bibinfo {author} {\bibfnamefont {J.}~\bibnamefont {Toner}},\ }\bibfield
  {title} {\bibinfo {title} {Universality class for a nonequilibrium state of
  matter: {A} $d = 4 -\epsilon$ expansion study of
  {Malthusian} flocks},\ }\href {https://doi.org/10.1103/PhysRevE.102.022610}
  {\bibfield  {journal} {\bibinfo  {journal} {Physical Review E}\ }\textbf
  {\bibinfo {volume} {102}},\ \bibinfo {pages} {022610} (\bibinfo {year}
  {2020}{\natexlab{b}})}\BibitemShut {NoStop}%
\bibitem [{\citenamefont {Bassler}\ and\ \citenamefont
  {Schmittmann}(1994)}]{bassler_prl94}%
  \BibitemOpen
  \bibfield  {author} {\bibinfo {author} {\bibfnamefont {K.~E.}\ \bibnamefont
  {Bassler}}\ and\ \bibinfo {author} {\bibfnamefont {B.}~\bibnamefont
  {Schmittmann}},\ }\bibfield  {title} {\bibinfo {title} {Critical {Dynamics}
  of {Nonconserved} {Ising}-{Like} {Systems}},\ }\href
  {https://doi.org/10.1103/PhysRevLett.73.3343} {\bibfield  {journal} {\bibinfo
   {journal} {Physical Review Letters}\ }\textbf {\bibinfo {volume} {73}},\
  \bibinfo {pages} {3343} (\bibinfo {year} {1994})}\BibitemShut {NoStop}%
\bibitem [{\citenamefont {Wong}\ and\ \citenamefont {Lee}(2025)}]{wong_a25}%
  \BibitemOpen
  \bibfield  {author} {\bibinfo {author} {\bibfnamefont {M.}~\bibnamefont
  {Wong}}\ and\ \bibinfo {author} {\bibfnamefont {C.~F.}\ \bibnamefont {Lee}},\
  }\href {https://doi.org/10.48550/arXiv.2507.06068} {\bibinfo {title} {New
  universality classes govern the critical and multicritical behavior of an
  active {Ising} model}} (\bibinfo {year} {2025}),\ \bibinfo {note}
  {arXiv:2507.06068 [cond-mat]}\BibitemShut {NoStop}%
\bibitem [{\citenamefont {Graner}\ and\ \citenamefont
  {Glazier}(1992)}]{graner_prl92}%
  \BibitemOpen
  \bibfield  {author} {\bibinfo {author} {\bibfnamefont {F.}~\bibnamefont
  {Graner}}\ and\ \bibinfo {author} {\bibfnamefont {J.~A.}\ \bibnamefont
  {Glazier}},\ }\bibfield  {title} {\bibinfo {title} {Simulation of biological
  cell sorting using a two-dimensional extended {Potts} model},\ }\href
  {https://doi.org/10.1103/PhysRevLett.69.2013} {\bibfield  {journal} {\bibinfo
   {journal} {Physical Review Letters}\ }\textbf {\bibinfo {volume} {69}},\
  \bibinfo {pages} {2013} (\bibinfo {year} {1992})}\BibitemShut {NoStop}%
\bibitem [{\citenamefont {Glazier}\ and\ \citenamefont
  {Graner}(1993)}]{glazier_pre93}%
  \BibitemOpen
  \bibfield  {author} {\bibinfo {author} {\bibfnamefont {J.~A.}\ \bibnamefont
  {Glazier}}\ and\ \bibinfo {author} {\bibfnamefont {F.}~\bibnamefont
  {Graner}},\ }\bibfield  {title} {\bibinfo {title} {Simulation of the
  differential adhesion driven rearrangement of biological cells},\ }\href
  {https://doi.org/10.1103/PhysRevE.47.2128} {\bibfield  {journal} {\bibinfo
  {journal} {Physical Review E}\ }\textbf {\bibinfo {volume} {47}},\ \bibinfo
  {pages} {2128} (\bibinfo {year} {1993})}\BibitemShut {NoStop}%
\bibitem [{\citenamefont {Wensink}\ \emph {et~al.}(2012)\citenamefont
  {Wensink}, \citenamefont {Dunkel}, \citenamefont {Heidenreich}, \citenamefont
  {Drescher}, \citenamefont {Goldstein}, \citenamefont {Löwen},\ and\
  \citenamefont {Yeomans}}]{wensink_pnas12}%
  \BibitemOpen
  \bibfield  {author} {\bibinfo {author} {\bibfnamefont {H.~H.}\ \bibnamefont
  {Wensink}}, \bibinfo {author} {\bibfnamefont {J.}~\bibnamefont {Dunkel}},
  \bibinfo {author} {\bibfnamefont {S.}~\bibnamefont {Heidenreich}}, \bibinfo
  {author} {\bibfnamefont {K.}~\bibnamefont {Drescher}}, \bibinfo {author}
  {\bibfnamefont {R.~E.}\ \bibnamefont {Goldstein}}, \bibinfo {author}
  {\bibfnamefont {H.}~\bibnamefont {Löwen}},\ and\ \bibinfo {author}
  {\bibfnamefont {J.~M.}\ \bibnamefont {Yeomans}},\ }\bibfield  {title}
  {\bibinfo {title} {Meso-scale turbulence in living fluids},\ }\href
  {https://doi.org/10.1073/pnas.1202032109} {\bibfield  {journal} {\bibinfo
  {journal} {Proceedings of the National Academy of Sciences}\ }\textbf
  {\bibinfo {volume} {109}},\ \bibinfo {pages} {14308} (\bibinfo {year}
  {2012})}\BibitemShut {NoStop}%
\bibitem [{\citenamefont {Katz}\ \emph {et~al.}(1983)\citenamefont {Katz},
  \citenamefont {Lebowitz},\ and\ \citenamefont {Spohn}}]{katz_prb83}%
  \BibitemOpen
  \bibfield  {author} {\bibinfo {author} {\bibfnamefont {S.}~\bibnamefont
  {Katz}}, \bibinfo {author} {\bibfnamefont {J.~L.}\ \bibnamefont {Lebowitz}},\
  and\ \bibinfo {author} {\bibfnamefont {H.}~\bibnamefont {Spohn}},\ }\bibfield
   {title} {\bibinfo {title} {Phase transitions in stationary nonequilibrium
  states of model lattice systems},\ }\href
  {https://doi.org/10.1103/PhysRevB.28.1655} {\bibfield  {journal} {\bibinfo
  {journal} {Physical Review B}\ }\textbf {\bibinfo {volume} {28}},\ \bibinfo
  {pages} {1655} (\bibinfo {year} {1983})}\BibitemShut {NoStop}%
\bibitem [{\citenamefont {Katz}\ \emph {et~al.}(1984)\citenamefont {Katz},
  \citenamefont {Lebowitz},\ and\ \citenamefont {Spohn}}]{katz_jstatphys84}%
  \BibitemOpen
  \bibfield  {author} {\bibinfo {author} {\bibfnamefont {S.}~\bibnamefont
  {Katz}}, \bibinfo {author} {\bibfnamefont {J.~L.}\ \bibnamefont {Lebowitz}},\
  and\ \bibinfo {author} {\bibfnamefont {H.}~\bibnamefont {Spohn}},\ }\bibfield
   {title} {\bibinfo {title} {Nonequilibrium steady states of stochastic
  lattice gas models of fast ionic conductors},\ }\href
  {https://doi.org/10.1007/BF01018556} {\bibfield  {journal} {\bibinfo
  {journal} {Journal of Statistical Physics}\ }\textbf {\bibinfo {volume}
  {34}},\ \bibinfo {pages} {497} (\bibinfo {year} {1984})}\BibitemShut
  {NoStop}%
\bibitem [{\citenamefont {Janssen}\ and\ \citenamefont
  {Schmittmann}(1986)}]{janssen_ZPB86}%
  \BibitemOpen
  \bibfield  {author} {\bibinfo {author} {\bibfnamefont {H.~K.}\ \bibnamefont
  {Janssen}}\ and\ \bibinfo {author} {\bibfnamefont {B.}~\bibnamefont
  {Schmittmann}},\ }\bibfield  {title} {\bibinfo {title} {Field theory of
  critical behaviour in driven diffusive systems},\ }\href
  {https://doi.org/10.1007/BF01312845} {\bibfield  {journal} {\bibinfo
  {journal} {Zeitschrift fuer Physik B Condensed Matter}\ }\textbf {\bibinfo
  {volume} {64}},\ \bibinfo {pages} {503} (\bibinfo {year} {1986})}\BibitemShut
  {NoStop}%
\bibitem [{\citenamefont {Leung}\ and\ \citenamefont
  {Cardy}(1986)}]{leung_jstatphys86}%
  \BibitemOpen
  \bibfield  {author} {\bibinfo {author} {\bibfnamefont {K.-t.}\ \bibnamefont
  {Leung}}\ and\ \bibinfo {author} {\bibfnamefont {J.~L.}\ \bibnamefont
  {Cardy}},\ }\bibfield  {title} {\bibinfo {title} {Field theory of critical
  behavior in a driven diffusive system},\ }\href
  {https://doi.org/10.1007/BF01011310} {\bibfield  {journal} {\bibinfo
  {journal} {Journal of Statistical Physics}\ }\textbf {\bibinfo {volume}
  {44}},\ \bibinfo {pages} {567} (\bibinfo {year} {1986})}\BibitemShut
  {NoStop}%
  \bibitem{kardar_b07}
  K.~Mehran, Statistical Physics of Fields (Cambridge University Press, 2012)
   \bibitem{kardar_prl86}
K.~Mehran, G.~Parisi, and Y.-C.~Zhang, Dynamic Scaling of Growing Interfaces,
\href
  {https://journals.aps.org/prl/abstract/10.1103/PhysRevLett.56.889} {\bibfield  {journal} {\bibinfo
  {journal} {Physical Review Letters}\ }\textbf {\bibinfo {volume}
  {56}},\ \bibinfo {pages} {889} (\bibinfo {year} {1986})}
\bibitem [{\citenamefont {Dupuis}\ \emph {et~al.}(2021)\citenamefont {Dupuis},
  \citenamefont {Canet}, \citenamefont {Eichhorn}, \citenamefont {Metzner},
  \citenamefont {Pawlowski}, \citenamefont {Tissier},\ and\ \citenamefont
  {Wschebor}}]{dupuis_physrep21}%
  \BibitemOpen
  \bibfield  {author} {\bibinfo {author} {\bibfnamefont {N.}~\bibnamefont
  {Dupuis}}, \bibinfo {author} {\bibfnamefont {L.}~\bibnamefont {Canet}},
  \bibinfo {author} {\bibfnamefont {A.}~\bibnamefont {Eichhorn}}, \bibinfo
  {author} {\bibfnamefont {W.}~\bibnamefont {Metzner}}, \bibinfo {author}
  {\bibfnamefont {J.}~\bibnamefont {Pawlowski}}, \bibinfo {author}
  {\bibfnamefont {M.}~\bibnamefont {Tissier}},\ and\ \bibinfo {author}
  {\bibfnamefont {N.}~\bibnamefont {Wschebor}},\ }\bibfield  {title} {\bibinfo
  {title} {The nonperturbative functional renormalization group and its
  applications},\ }\href {https://doi.org/10.1016/j.physrep.2021.01.001}
  {\bibfield  {journal} {\bibinfo  {journal} {Physics Reports}\ }\textbf
  {\bibinfo {volume} {910}},\ \bibinfo {pages} {1} (\bibinfo {year}
  {2021})}\BibitemShut {NoStop}%
  \bibitem{SM} Supplemental material.
\end{thebibliography}
\end{document}